\documentclass[twocolumn,twocolappendix]{aastex631}
\usepackage{amsmath}
\usepackage{graphicx}
\usepackage{upgreek}
\usepackage{enumerate}
\usepackage{natbib}
\usepackage{hyperref}

\newcommand{\M}{M_{\ast}}

\usepackage{color}

\usepackage{amsmath}

\received{XXX}
\revised{YYY}
\accepted{ZZZ}

\submitjournal{ApJ}

\shorttitle{ELG conformity}
\shortauthors{Gao et al.}

\begin{document}

\title{The DESI One-Percent Survey: A concise model for galactic conformity of ELGs}

\author{Hongyu Gao}
\affil{Department of Astronomy, School of Physics and Astronomy, Shanghai Jiao Tong University, Shanghai, 200240, People's Republic of China}

\author[0000-0002-4534-3125]{Y.P. Jing}
\affil{Department of Astronomy, School of Physics and Astronomy, Shanghai Jiao Tong University, Shanghai, 200240, People's Republic of China}
\affil{Tsung-Dao Lee Institute, and Shanghai Key Laboratory for Particle Physics and Cosmology, Shanghai Jiao Tong University, Shanghai, 200240, People's Republic of China}

\author[0000-0002-7697-3306]{Kun Xu}
\affil{Department of Astronomy, School of Physics and Astronomy, Shanghai Jiao Tong University, Shanghai, 200240, People's Republic of China}
\affil{Institute for Computational Cosmology, Department of Physics, Durham University, South Road, Durham DH1 3LE, UK}

\author{Donghai Zhao}
\affil{Department of Astronomy, School of Physics and Astronomy, Shanghai Jiao Tong University, Shanghai, 200240, People's Republic of China}
\affil{Key Laboratory for Research in Galaxies and Cosmology, Shanghai Astronomical Observatory, Shanghai, 200030, People's Republic of China}

\author{Shanquan Gui}
\affil{Department of Astronomy, School of Physics and Astronomy, Shanghai Jiao Tong University, Shanghai, 200240, People's Republic of China}

\author[0000-0001-6575-0142]{Yun Zheng}
\affiliation{Department of Astronomy, School of Physics and Astronomy, Shanghai Jiao Tong University, Shanghai, 200240, People's Republic of China}

\author{Xiaolin Luo}
\affil{Department of Astronomy, School of Physics and Astronomy, Shanghai Jiao Tong University, Shanghai, 200240, People's Republic of China}

\author{Jessica Nicole Aguilar}
\affil{Lawrence Berkeley National Laboratory, 1 Cyclotron Road, Berkeley, CA 94720, USA}

\author[0000-0001-6098-7247]{Steven Ahlen}
\affil{Physics Dept., Boston University, 590 Commonwealth Avenue, Boston, MA 02215, USA}

\author{David Brooks}
\affil{Department of Physics \& Astronomy, University College London, Gower Street, London, WC1E 6BT, UK}

\author{Todd Claybaugh}
\affil{Lawrence Berkeley National Laboratory, 1 Cyclotron Road, Berkeley, CA 94720, USA}

\author[0000-0002-5954-7903]{Shaun Cole}
\affil{Institute for Computational Cosmology, Department of Physics, Durham University, South Road, Durham DH1 3LE, UK}

\author[0000-0002-1769-1640]{Axel de la Macorra}
\affil{Instituto de F\'{\i}sica, Universidad Nacional Aut\'{o}noma de M\'{e}xico, Cd. de M\'{e}xico C.P. 04510, M\'{e}xico}

\author[0000-0002-2890-3725]{Jaime E. Forero-Romero}
\affil{Departamento de F\'isica, Universidad de los Andes, Cra. 1 No. 18A-10, Edificio Ip, CP 111711, Bogot\'a, Colombia}
\affil{Observatorio Astron\'omico, Universidad de los Andes, Cra. 1 No. 18A-10, Edificio H, CP 111711 Bogot\'a, Colombia}

\author[0000-0003-3142-233X]{Satya Gontcho A Gontcho}
\affil{Lawrence Berkeley National Laboratory, 1 Cyclotron Road, Berkeley, CA 94720, USA}

\author[0000-0002-6024-466X]{Mustapha Ishak}
\affil{Department of Physics, The University of Texas at Dallas, Richardson, TX 75080, USA}

\author{Andrew Lambert}
\affil{Lawrence Berkeley National Laboratory, 1 Cyclotron Road, Berkeley, CA 94720, USA}

\author[0000-0003-1838-8528]{Martin Landriau}
\affil{Lawrence Berkeley National Laboratory, 1 Cyclotron Road, Berkeley, CA 94720, USA}

\author[0000-0003-4962-8934]{Marc Manera}
\affil{Institut de F\'{i}sica d’Altes Energies (IFAE), The Barcelona Institute of Science and Technology, Campus UAB, 08193 Bellaterra Barcelona, Spain}
\affil{Departament de F\'{i}sica, Serra H\'{u}nter, Universitat Aut\`{o}noma de Barcelona, 08193 Bellaterra (Barcelona), Spain}

\author[0000-0002-1125-7384]{Aaron Meisner}
\affil{NSF's NOIRLab, 950 N. Cherry Ave., Tucson, AZ 85719, USA}

\author{Ramon Miquel}
\affil{Institut de F\'{i}sica d’Altes Energies (IFAE), The Barcelona Institute of Science and Technology, Campus UAB, 08193 Bellaterra Barcelona, Spain}
\affil{Instituci\'{o} Catalana de Recerca i Estudis Avan\c{c}ats, Passeig de Llu\'{\i}s Companys, 23, 08010 Barcelona, Spain}

\author[0000-0001-6590-8122]{Jundan Nie}
\affil{National Astronomical Observatories, Chinese Academy of Sciences, A20 Datun Rd., Chaoyang District, Beijing, 100012, P.R. China}

\author[0000-0001-5589-7116]{Mehdi Rezaie}
\affil{Department of Physics, Kansas State University, 116 Cardwell Hall, Manhattan, KS 66506, USA}

\author{Graziano Rossi}
\affil{Department of Physics and Astronomy, Sejong University, Seoul, 143-747, Korea}

\author[0000-0002-9646-8198]{Eusebio Sanchez}
\affil{CIEMAT, Avenida Complutense 40, E-28040 Madrid, Spain}

\author{Michael Schubnell}
\affil{Department of Physics, University of Michigan, Ann Arbor, MI 48109, USA}

\author[0000-0002-6588-3508]{Hee-Jong Seo}
\affil{Department of Physics \& Astronomy, Ohio University, Athens, OH 45701, USA}

\author[0000-0003-1704-0781]{Gregory Tarl\'{e}}
\affil{University of Michigan, Ann Arbor, MI 48109, USA}

\author{Benjamin Alan Weaver}
\affil{NSF's NOIRLab, 950 N. Cherry Ave., Tucson, AZ 85719, USA}

\author[0000-0002-4135-0977]{Zhimin Zhou}
\affil{National Astronomical Observatories, Chinese Academy of Sciences, A20 Datun Rd., Chaoyang District, Beijing, 100012, P.R. China}

\correspondingauthor{Y.P. Jing}
\email{ypjing@sjtu.edu.cn}

\begin{abstract}
Galactic conformity is the phenomenon in which a galaxy of a certain physical property is correlated with its neighbors of the same property, implying a possible causal relationship. The observed auto correlations of emission line galaxies (ELGs) from the highly complete DESI One-Percent survey exhibit a strong clustering signal on small scales, providing clear evidence for the conformity effect of ELGs. Building upon the original subhalo abundance matching (SHAM) method developed by Gao et al. (2022, 2023), we propose a concise conformity model to improve the ELG-halo connection. In this model, the number of satellite ELGs is boosted by a factor of $\sim 5$ in the halos whose central galaxies are ELGs. We show that the mean ELG satellite number in such central halos is still smaller than 1, and the model does not significantly increase the overall satellite fraction. With this model, we can well recover the ELG auto correlations to the smallest scales explored with the current data (i.e. $r_{\mathrm{p}} > 0.03$ $\mathrm{Mpc}\,h^{-1}$ in real space and at $s > 0.3$ $\mathrm{Mpc}\,h^{-1}$ in redshift space), while the cross correlations between luminous red galaxies (LRGs) and ELGs are nearly unchanged. Although our SHAM model has only 8 parameters, we further verify that it can accurately describe the ELG clustering in the entire redshift range from $z = 0.8$ to $1.6$. We therefore expect that this method can be used to generate high-quality ELG lightcone mocks for DESI.
\end{abstract}


\section{Introduction} \label{sec:intro}

In the standard cosmological model, dark matter halos can accrete surrounding gas and foster the formation of galaxies (\citealt{1991ApJ...379...52W}; see more references in
\citealt{2018ARA&A..56..435W}). Emission line galaxies (ELGs) have been employed as one of the primary targets for current dark energy surveys \citep{2016arXiv161100036D, 2014PASJ...66R...1T}. ELGs are mainly late-type galaxies with significant ongoing star-forming activity, and thus their galaxy-halo connection may differ from that of normal galaxies. Several studies on halo occupation distribution \citep[HOD,][]{2012MNRAS.426..679G, 2013MNRAS.432.2717C, 2020MNRAS.499.5486A, 2020MNRAS.497..581A, 2021MNRAS.505.2784Z, 2021PASJ...73.1186O, 2022MNRAS.510.3301Y, abacusELG_Rocher}, conditional stellar mass function \citep[CSMF,][]{2019ApJ...871..147G}, subhalo abundance matching \citep[SHAM,][]{2016MNRAS.461.3421F, 2017MNRAS.472..550F, 2022ApJ...928...10G, Gao, 2023MNRAS.519.4253L, inclusiveSHAM, overviewSHAM}, semi-analytical model \citep[SAM,][]{2018MNRAS.474.4024G, 2020MNRAS.498.1852G} and hydrodynamical simulation \citep{2021MNRAS.502.3599H, 2022MNRAS.512.5793Y, 2022arXiv221010068H, 2022arXiv221010072H} suggest that ELGs tend to reside in host halos with masses $\sim 10^{12}$ $M_{\odot}\, h^{-1}$ and the overall satellite fraction of ELGs is less than 20\%. Due to various parameterized assumptions and target selections of ELGs in different surveys with specific color cuts and photometric depths, the ELG-halo connection models vary widely. And the accuracy of these models also varies widely in describing the clustering of ELGs, especially on small scales.

In most previous models of the ELG-halo connection, the central and its subhalos are usually assumed to be uncorrelated in forming ELGs. For example, by analyzing the auto and cross correlations of ELGs and luminous red galaxies (LRGs) in the DESI One-Percent survey \citep{sv}, \cite{Gao} improved the SHAM method \citep{2022ApJ...928...10G} to construct the ELG-halo connection. This method shows an accurate modeling of auto and cross correlations of ELGs and LRGs, except that it still underestimates the strong one-halo term of the ELG auto correlation at $r_{\mathrm{p}}<0.3$ $\mathrm{Mpc}\,h^{-1}$ ($s < 1$ $\mathrm{Mpc}\,h^{-1}$). For the same survey, \cite{inclusiveSHAM} and \cite{overviewSHAM} also adopted different SHAM methods to model the ELG clustering in redshift space, although both of their models fit the observations only at $s>5$ $\mathrm{Mpc}\,h^{-1}$. 

Utilizing the highly complete ELG sample in the One-Percent survey, a series of analysis have shown a strong clustering signal of ELGs within the scale of $\sim 0.5$ $\mathrm{Mpc}\,h^{-1}$ \citep{Gao, abacusELG_Rocher}. In fact, the prominent clustering of ELGs has been found in the surveys prior to DESI. For instance, \cite{2022ApJ...928...10G}  noticed that the clustering of the ELGs with the strongest $\mathrm{[O\,II]}$ exhibits an excess trend in the one-halo term. The angular correlation of ELGs at $z > 1$ measured by \cite{2021PASJ...73.1186O} also presents a clear strong signal at $\sim 3.6$ $\mathrm {arcsec}$. The high abundance of galaxy pairs at small scales may indicate a correlation between the close ELG neighbors in their physical properties. This effect is known as the galactic conformity \citep[e.g.,][]{2006MNRAS.366....2W, 2013MNRAS.430.1447K, 2015MNRAS.452.1958H, 2017MNRAS.470.1298P, 2018MNRAS.475.1177L, 2018MNRAS.476.1637Z, 2018MNRAS.477.3136S, 2022MNRAS.511.1789Z}. Using the group catalog constructed by \cite{2007ApJ...671..153Y}, \cite{2006MNRAS.366....2W} discovered that for a fixed halo mass, the fraction of late (early)-type satellites is significantly higher in the halo hosting the same type central galaxy (i.e. one-halo galactic conformity). Beyond the halo scale, \cite{2013MNRAS.430.1447K} showed that the conformity effect of low-mass central galaxies ($<10^{10.0}M_{\odot}$) can affect their neighbors more than 4 $\mathrm{Mpc}$ away (i.e. the two-halo galactic conformity), although some later studies \citep[e.g.,][]{2017MNRAS.471.1192S, 2018MNRAS.477..935T} argued that the two-halo conformity may be due to artificial effects in their sample selection.

It is difficult to interpret the high clustering of ELGs on very small scales with usual SHAM or HOD models. In the SHAM model of \cite{Gao} (hereafter Paper I), they have found that ELGs can be randomly selected from a general population of galaxies, but the satellite fraction of ELGs in massive halos needs to be reduced (also a kind of galactic conformity). Even though the model has retained the satellite ELGs in the small halos as much as possible, the modeled clustering of ELGs is still not strong enough to match the observed one. This strong clustering is therefore considered as evidence for the one-halo conformity effect of ELG. Meanwhile, \cite{abacusELG_Rocher} combined a conformity model with different HODs to recover the ELG auto correlations down to small scales of $\sim 0.04$ $\mathrm{Mpc}\,h^{-1}$. In this model, satellite ELGs can only exist in halos where the central galaxy is an ELG. By the design, the model would predict a very weak (even negative) cross correlation between ELGs and LRGs on small scales within the halo size of LRGs, which is not observed as we will see below. 

In this paper, we aim to carefully investigate the ELG conformity effect and develop a model to explain it. As will be shown shortly, using a concise model with {\textit{one}} parameter, we can incorporate this effect into the original SHAM method in Paper I. The improved model is able to effectively reproduce the observed clustering of all ELG samples at $0.8<z<1.6$ down to the smallest scales that can be measured with the current data [i.e. at $r_{\mathrm{p}} > 0.03$ ($s > 0.3$) $\mathrm{Mpc}\,h^{-1}$ in real (redshift) space], and is applicable to the entire redshift range that the DESI ELG survey spans. 

The paper is structured as follows. In Section \ref{sec:data}, we briefly describe the data and the clustering measurements. In Section \ref{sec:method}, we descibe how we improve the original SHAM method by introducing a conformity model. We then show the results in Section \ref{sec:Results}. We finally make a conclusion in Section \ref{sec:Conclusion}. The cosmological parameters adopted in our work are $\Omega_{\mathrm{m},0} = 0.268$, $\Omega_{\Lambda,0} = 0.732$ and $H_0 = 100h \,\mathrm{km\, s^{-1}\,Mpc^{-1}}=71 \,\mathrm{km\,s^{-1}\,Mpc^{-1}}$.

There are also several parallel studies focusing on the connection between galaxies and halos in the DESI One-Percent survey \citep{abacusLRGQSO_Yuan, abacusELG_Rocher, inclusiveSHAM, overviewSHAM, 2023arXiv231009329Y}. 

\section{Data and Simulation} \label{sec:data}

\subsection{DESI One-Percent survey and galaxy sample} \label{One-Percent survey}

DESI is a Stage IV dark energy survey that aims to collect the spectra of about 40 million extragalactic objects over five years \citep{2013arXiv1308.0847L, 2016arXiv161100036D,2016arXiv161100037D, 2022AJ....164..207D}. This survey will cover more than 14,000 $\mathrm{deg^2}$ in sky, and has been conducted using a multi-object fiber-fed spectrograph mounted on the prime focus panel of the 4-meter Mayall Telescope situated in Kitt Peak National Observatory \citep{2022AJ....164..207D}. The spectrometer has a wavelength range of $3600-9800\, \AA$ and can allocate fibers to 5,000 objects at one visit \citep{2016arXiv161100037D, 2023AJ....165....9S, corrector}. Several auxiliary pipelines supporting the DESI experiment are described in \cite{2023AJ....165..144G, redrock2023, fba, ops, 2023AJ....165...50M}. The strategy of DESI target selection and survey validation have been presented in a series of works \citep{2020RNAAS...4..188A, 2020RNAAS...4..187R, 2020RNAAS...4..181Z, 2020RNAAS...4..180R, 2020RNAAS...4..179Y, 2023ApJ...943...68L, 2023AJ....165..124A, 2023ApJ...947...37C, 2023AJ....165..253H, 2023ApJ...944..107C, 2023AJ....165..126R, 2023AJ....165...58Z}. 

As the third phase of the survey validation \citep[SV3,][]{sv}, the One-Percent survey is used to further optimize and verify the efficiency of the observation and manipulation procedures. Its footprint covers twenty discrete small regions of the sky, each covering an area of about 7 $\mathrm{deg^2}$. The One-Percent survey contains a total of 488 tiles, which consists of $\sim 10-11$ ($\sim 12-13$) repeated visits in bright (dark) time for each field. As the result, the efficiency of fiber placement and the success rate of spectroscopic measurements are quite high. More than 99\% of LRGs and 95\% of ELG samples can be successfully assigned to fibers. Consequently, galaxy samples in the One-Percent survey suffer little from fiber collisions, allowing the measurement of clustering to be extended to smaller scales ($< 0.1 \mathrm{Mpc} \, h^{-1}$).

The DESI Early Data Release \citep[EDR,][]{dr} provides both the full and the clustering catalogs for the One-Percent survey. We take the ELG and LRG samples with successful redshift measurements from the clustering catalog. For each galaxy, with five-band photometry $grzW1W2$ \citep{2017PASP..129f4101Z,2019AJ....157..168D, dr9,  2005astro.ph.10346T, 2010AJ....140.1868W}, we conduct a spectral energy distribution (SED) fit using {\tt\string CIGALE} \citep{2019A&A...622A.103B}. See Paper I for detailed template and model settings. Using exactly the same approach as in Paper I, we select 4 ELG subsamples (ELG0, ELG1, ELG2, ELG3) and 3 LRG subsamples (LRG0, LRG1, LRG2) binned by stellar mass. We measure the auto correlation functions of the different subsamples and the cross correlations between the ELG and LRG subsamples. In addition, we also select the entire ELG sample (without stellar mass bins) at different redshifts in our analysis. The details of our galaxy samples are shown in Table \ref{tab:elglrg}. In the measurements of the entire ELG samples, we correct the fiber collision by combining the completeness weight derived from the 128 Merged Target List (MTL) realizations \citep{ELG_Lasker}, the pairwise-inverse-probability (PIP) weight \citep{2017MNRAS.472.1106B} and the angular-up weight (ANG) \citep{2017MNRAS.472L..40P, 2020MNRAS.498..128M}. The combination of these weights can increase the correlation function by $\sim 10\%$ at $\sim 0.1\mathrm{Mpc} \, h^{-1}$.		

\begin{deluxetable}{cccc}
	\tablenum{1}
	\tablecaption{Details of the galaxy samples used in our analysis. }
	\label{tab:elglrg}
	\tablehead{ \colhead{Name} &
		\colhead{Redshift Range} & \colhead{$\log \M\, [M_{\odot}]$ } &\colhead{$N_{\mathrm{g}}$}
	}
	\startdata
	LRG0&$0.8<z\leq 1.0$& $\left[11.1, 11.3\right]$& 13906 \\
	LRG1&$0.8<z\leq 1.0$& $\left[11.3, 11.5\right]$& 4834 \\
	LRG2&$0.8<z\leq 1.0$& $\left[11.5, 11.7\right]$& 957 \\
	ELG0&$0.8<z\leq 1.0$& $\left[8.5, 9.0\right]$& 9481 \\
    ELG1&$0.8<z\leq 1.0$& $\left[9.0, 9.5\right]$& 29764 \\
    ELG2&$0.8<z\leq 1.0$& $\left[9.5, 10.0\right]$& 34155 \\
    ELG3&$0.8<z\leq 1.0$& $\left[10.0, 10.5\right]$& 6583 \\
	All ELG&$0.8<z\leq 1.0$& \nodata&  82887 \\
    All ELG&$1.0<z\leq 1.2$& \nodata&  68366 \\
    All ELG&$1.2<z\leq 1.4$& \nodata&  54783 \\
    All ELG&$1.4<z\leq 1.6$& \nodata&  37756 \\
	\enddata
	%\tablecomments{}
\end{deluxetable}

\subsection{Clustering measurements} \label{clustering}

We adopt the classic Landy-Szalay estimator \citep{1993ApJ...412...64L, 1998ApJ...494L..41S} to estimate the two-point correlation function of subsamples $x$ and $y$:
\begin{equation}
	\xi_{xy}\left(r_{\mathrm{p}}, r_{\mathrm{\pi}}\right) = \left[ \frac{D_xD_y-D_xR_y-D_yR_x+R_xR_y}{R_xR_y}\right]\left(r_{\mathrm{p}}, r_{\mathrm{\pi}}\right),
\end{equation}
where $D_xD_y$, $D_xR_y$, $D_yR_x$ and $R_xR_y$ denote the normalized weighted pair counts in each $\left(r_{\mathrm{p}}, r_{\mathrm{\pi}}\right)$ bin. In the measurement, we set 25 $r_{\mathrm{p}}$ bins from $10^{-1.5}$ to $10^{1.477}$ $\mathrm{Mpc}\,h^{-1}$ in logarithmic space and 40 $r_{\mathrm{\pi}}$ bins from $0$ to $40$ $\mathrm{Mpc}\,h^{-1}$ in linear space.

We then integrate the $\xi_{xy}\left(r_{\mathrm{p}}, r_{\mathrm{\pi}}\right)$ along the direction of $r_{\mathrm{\pi}}$ \citep{1983ApJ...267..465D}:
\begin{equation}
	w_{\mathrm{p},xy}\left(r_{\rm p}\right) = 2\int_{0}^{r_{\pi, \rm max}} \xi_{xy} \left( r_{\rm p}, r_{\pi} \right)\mathrm{d}r_{\pi},
\end{equation}
where $w_{\mathrm{p},xy} \left(r_{\mathrm{p}}\right)$ is the projected correlation function in real space and $r_{\pi, \rm max}=40 \,\mathrm{Mpc}\,h^{-1}$ is the upper limit of the integration.

Moveover, the multipole moments in redshift space can be calculated through \citep{1992ApJ...385L...5H}
\begin{equation}
	\xi_{l,xy}\left(s\right)=\frac{2l+1}{2}\int_{-1}^{1}\xi_{xy}\left(s,\mu\right)L_l\left(\mu\right) \mathrm{d}\mu,
\end{equation}
where $s = \sqrt{r^2_{\mathrm{p}} + r^2_{\mathrm{\pi}}}$ and $\mu = r_{\mathrm{\pi}}/s$. The monopole, quadrupole and hexadecapole corresponds to $l=0$, $2$ and $4$, respectively. We adopt 15 $s$ bins from $0.3$ to $30$ $\mathrm{Mpc}\,h^{-1}$ in logarithmic space and 10 $\mu$ bins from $0$ to $1$ in linear space.

We divide the area of the One-Percent survey into 100 approximately equal regions and estimate the covariance matrix of $\boldsymbol{w}_{\mathrm{p}}$ and $\boldsymbol{\xi}_{l}$ using a jackknife technique.

\subsection{N-body simulation} \label{simulation}
 
The N-body simulation used in this study is from {\tt\string CosmicGrowth} \citep{2019SCPMA..6219511J}. The boxsize of this simulation is $600$ $\mathrm{Mpc}\,h^{-1}$ and the number of particles is $3072^3$. The simulation sets the standard $\Lambda$CDM cosmological parameters: $\Omega_{\mathrm{m},0} = 0.268$, $\Omega_{\Lambda,0} = 0.732$, $h=0.71$, $n_{\mathrm{s}}=0.968$ and $\sigma_8=0.83$ and was run with an adaptive P$^3$M algorithm \citep{2002ApJ...574..538J}. The achieved mass resolution is $m_{\mathrm{p}} = 5.54 \times 10^8 \,M_{\odot}\,h^{-1}$.

The halo catalog is identified using the friends-of-friends (FOF) algorithm \citep{1985ApJ...292..371D}. We define the virial mass $M_{\mathrm{vir}} = \frac{4}{3} \pi R^3_{\mathrm{vir}} \Delta_{\mathrm{vir}}$ as the default halo mass $M_{\mathrm{h}}$, where $\Delta_{\mathrm{vir}}$ is the over-density for a virialized spherical structure \citep{1998ApJ...495...80B}. The subhalo catalog and merger trees are constructed through the Hierarchical-Bound-Tracing algorithm ({\tt\string HBT+}) \citep{2012MNRAS.427.2437H,2018MNRAS.474..604H}. We define the default mass $M_{\mathrm{s}}$ of the subhalo as the virial mass at the last moment before infall. We further calculate the merging time scale of the subhalos with particle number less than 20 to determine whether they are still alive \citep{2008ApJ...675.1095J}. The final halo and subhalo mass functions are consistent with the theoretical expectations \citep{2019SCPMA..6219511J, 2022ApJ...925...31X}.

Same as Paper I, the redshift space distortion (RSD), is added along the $z$-direction for each halo and subhalo. When modeling the galaxy clustering in redshift space, we assume a Gaussian scatter $\sigma_z = 10$ $\mathrm{km\,s^{-1}}$ for ELGs and $\sigma_z = 40$ $\mathrm{km\,s^{-1}}$ for LRGs to account for the redshift uncertainties. Besides, we also incorporate the velocity bias \citep[e.g.,][]{2003ApJ...590..654Y} for the central galaxy using a random Gaussian distribution with $\sigma_{\mathrm{c}} = \alpha_{\mathrm{c}} \times \sigma_v$, where $\sigma_v = \sqrt{GM_{\mathrm{vir}}/\left( 2 R_{\mathrm{vir}} \right)}$ is the one-dimensional velocity dispersion of a halo and $\alpha_{\mathrm{c}}$ is fixed to $0.22$ \citep{2015MNRAS.446..578G}.

We take four snapshots at $z=0.92$, $1.09$, $1.27$ and $1.47$ to cover the redshift range of $0.8<z<1.6$ in DESI.

\section{ELG-halo connection} \label{sec:method}

\begin{figure*}
	\centering
	\includegraphics[scale=0.6]{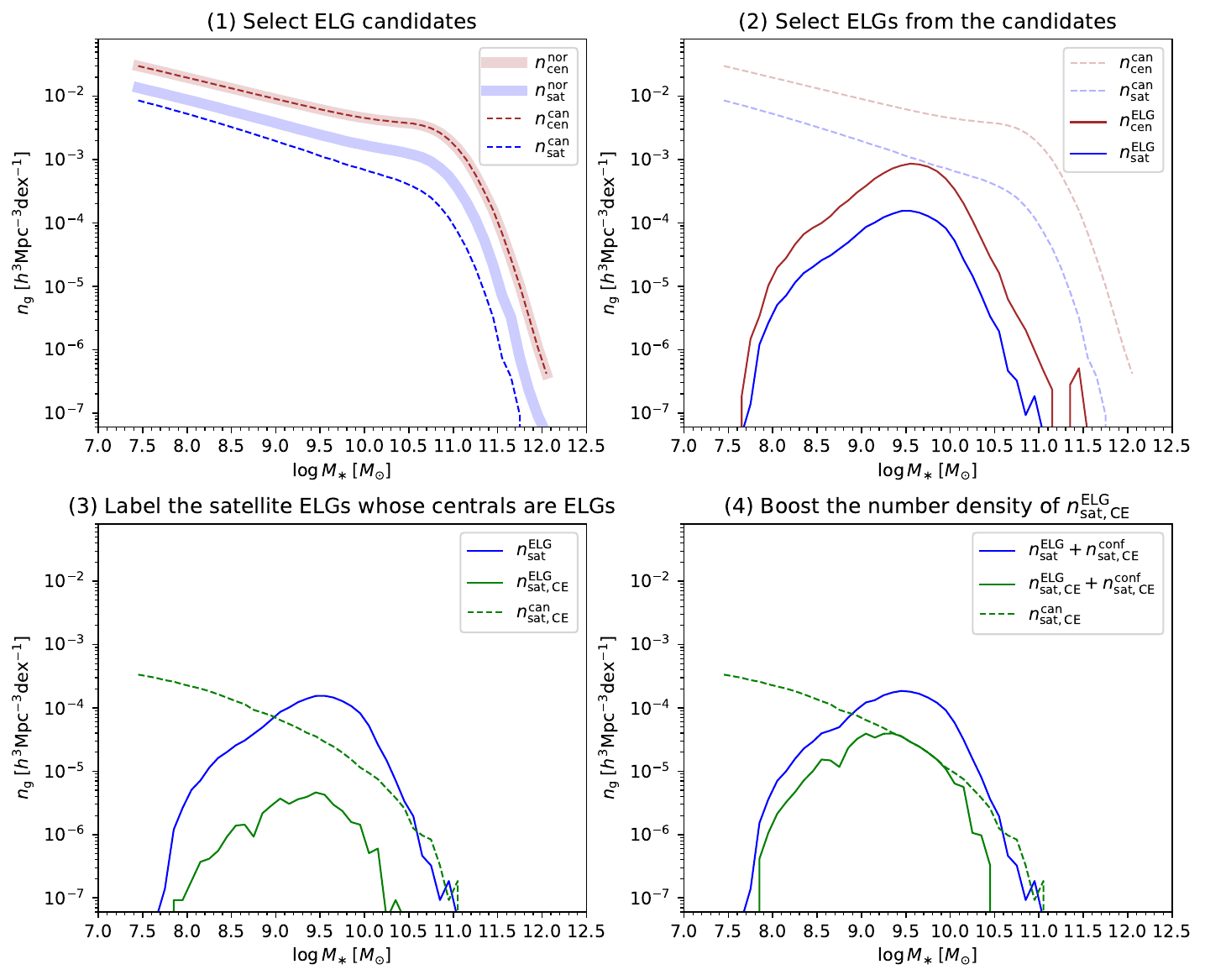}
    \caption{Illustration of the entire process of modeling ELG in simulation. The process consists of four steps, and we use the stellar mass functions (SMFs) to show various galaxy populations selected at each step. (1) {\it Top left}: populating halos (subhalos) with normal galaxies ($n^{\mathrm{nor}}_{\mathrm{cen}}$ and $n^{\mathrm{nor}}_{\mathrm{sat}}$). From the normal galaxies, selecting ELG candidates ($n^{\mathrm{can}}_{\mathrm{cen}}$ and $n^{\mathrm{can}}_{\mathrm{sat}}$) that have the same clustering as the observed. (2) {\it Top right}: selecting ELGs from the candidates according to the number observed in the DESI survey ($n^{\mathrm{ELG}}_{\mathrm{cen}}$ and $n^{\mathrm{ELG}}_{\mathrm{sat}}$). (3) {\it Bottom left}: labeling the satellite ELGs and satellite candidates whose central galaxies are ELGs as $n^{\mathrm{ELG}}_{\mathrm{sat},\mathrm{CE}}$ and $n^{\mathrm{can}}_{\mathrm{sat},\mathrm{CE}}$ respectively. (4) {\it Bottom right}: boosting the number of satellite ELGs around central ELGs. We use $n^{\mathrm{conf}}_{\mathrm{sat},\mathrm{CE}}$ to indicate the boosted ELGs due to the conformity effect. The first two steps represent the original SHAM method without conformity. The last two steps denote the modeling of the conformity effect. Here the conformity parameter $K_{\mathrm{conf}}$ is set to 10 for illustration only (see Section \ref{sec:AM_conformity} for more details). }
	%\caption{Illustration of the entire process of modeling ELG in simulation. The process consists of four steps: (1) populating halos (subhalos) with normal galaxies and selecting ELG candidates from them ({\it top left}); (2) selecting ELGs from the candidates ({\it top right}); (3) labeling the number density of those satellite ELGs and satellite candidates whose central galaxies are ELGs as $n^{\mathrm{ELG}}_{\mathrm{sat},\mathrm{CE}}$ and $n^{\mathrm{can}}_{\mathrm{sat},\mathrm{CE}}$ ({\it bottom left}); (4) boosting $n^{\mathrm{ELG}}_{\mathrm{sat},\mathrm{CE}}$ to increase the number of satellite ELGs around central ELGs ({\it bottom right}). The first two steps represent the original SHAM method without conformity. The last two steps denote the modeling of the conformity effect. Here the conformity parameter $K_{\mathrm{conf}}$ is set to 10 for illustration only (see Section \ref{sec:AM_conformity} for more details). In each panel, we show the stellar mass functions of various galaxy populations involved in this step with different styles of lines.}
	\label{fig:number_boost}
\end{figure*}

\subsection{Subhalo abundance matching without conformity} \label{sec:AM}
The basic idea of the original SHAM method developed by \cite{2022ApJ...928...10G} and Paper I is to first obtain normal galaxies in dark matter halos, and then to reduce the satellite fraction in massive halos to get ELG candidates. The final ELG sample can be randomly selected from the candidates by matching the observed stellar mass function (SMF) of ELG.
\subsubsection{Populate halo (subhalo) with normal galaxies} 
Given a halo or subhalo in the N-body simulation, we can assign a stellar mass $M_{\ast}$ to it through the stellar-halo mass relation (SHMR) model:
\begin{equation}
p(\M|M_{\mathrm{h}}) = \frac{1}{\sqrt{2\pi}\sigma}\exp\left[-\frac{\left(\log \M - \log \left\langle \M|M_{\mathrm{h}} \right\rangle\right)^2}{2\sigma^2}\right], \label{equ:p(M_star|M_h)}
\end{equation}
where $M_{\mathrm{h}}$ is the mass of the halo (subhlao), and the intrinsic dispersion of the SHMR is quantified by a Gaussian scatter $\sigma$. The mean relation of $M_{\ast}$ and $M_{\mathrm{h}}$ can be formulated by a double power-law function \citep{2006MNRAS.371..537W,2010MNRAS.402.1796W,2012ApJ...752...41Y,2013MNRAS.428.3121M}:
\begin{equation}
\left\langle \M|M_{\mathrm{h}} \right\rangle = \frac{2k}{\left(M_{\mathrm{h}}/M_0\right)^{-\alpha} + \left(M_{\mathrm{h}}/M_0\right)^{-\beta}} 
\end{equation}
where the parameters $\alpha$ and $\beta$ denote the slopes at the high and low-mass ends respectively, $M_0$ is the dividing point of the two power-laws, and the constant $k$ is used for normalization. 

Once the SHMR is applied to all the halos and subhalos, we can generate galaxies with stellar masses in the simulation. Using the best-fit SHMR (see the first five columns of Table \ref{tab:parameters}) provided by Paper I, we place galaxies in each halo and subhalo in the simulation. The number densities of these central  $n^{\mathrm{nor}}_{\mathrm{cen}}$ and satellite galaxies $n^{\mathrm{nor}}_{\mathrm{sat}}$ are shown as the shaded lines in the top left panel of Figure \ref{fig:number_boost}. These galaxies are complete in terms of stellar mass, free from any selection effects, and therefore are called normal galaxies. As shown by the SMFs in Paper I, the LRG and ELG samples in DESI are subject to complex selection effects. LRGs are predominantly massive red galaxies and can be considered nearly complete only at the massive end ($> 10^{11.3}$ $M_{\odot}$). ELGs make up only a small fraction ($< 10\%$) of the total normal galaxy population at each stellar mass, and are biased against satellite galaxies in massive halos. In the next subsection, we will mimic the selection functions of LRGs and ELGs and select them from the normal galaxies.

\begin{figure*}
	\centering
	\includegraphics[scale=0.6]{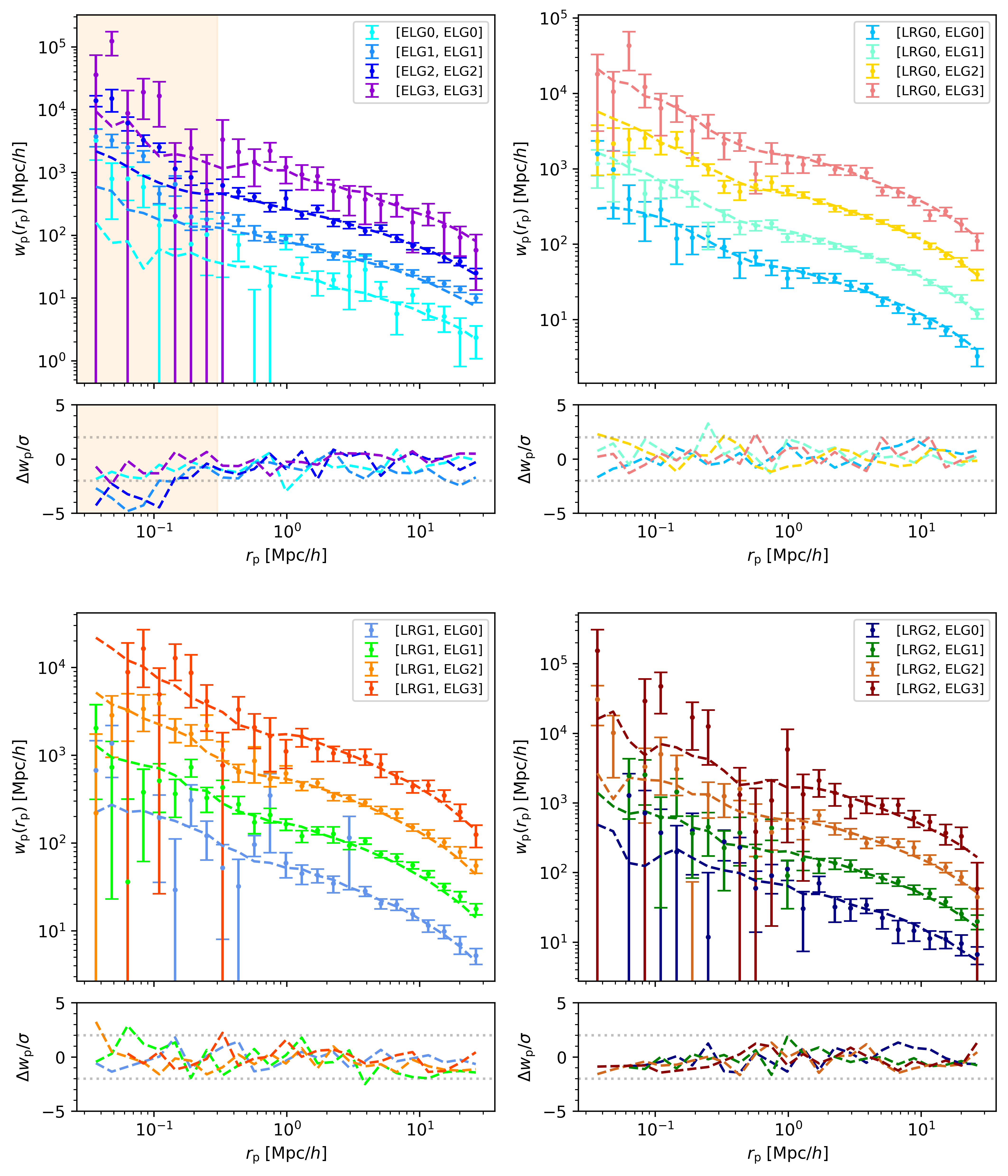}
	\caption{Comparison between the observed projected correlation functions at $0.8<z\leq 1.0$ and the model predictions without conformity. The top left panel shows the ELG auto correlations, and the other three panels display the LRGxELG cross correlations. The observations are represented as the data points with error bars, while the dashed lines are the model predictions. The differences between model and data, scaled by the measurement errors, are also plotted at the bottom of each panel. For the purpose of clarity, we multiply each $\boldsymbol{w}_{\mathrm{p}}$ by a factor of $3^n$ where $n=0,1,2,3$ correspond to the four ELG subsamples $0,1,2,3$ respectively. As shown in the top left panel, the models clearly underestimate the observed ELG auto correlations within $r_{\mathrm{p}} \sim 0.3$ $\mathrm{Mpc}\,h^{-1}$ (yellow shaded area). This means that the original model cannot provide enough galaxy pairs on small scales if the conformity effect is not taken into account.}
	\label{fig:wp_noconformity}
\end{figure*}

%\newpage
\subsubsection{Select ELGs from normal galaxies} \label{sec:two_step}
Next, we can select ELGs from the normal galaxies. The selection can be further divided into two steps. 

The first step is to select ELG candidates. As demonstrated in Paper I, relative to a central galaxy, we need to reduce the probability $P_{\mathrm{sat}}$ that a satellite galaxy can be selected as an ELG candidate. The physical motivation for this hypothesis is that star formation of satellite galaxies may be quenched by the dense environment, reducing the fraction of satellite ELGs. In Paper I, we proposed a halo mass-dependent model for $P_{\mathrm{sat}}$:
\begin{equation}
\begin{aligned}
P_{\mathrm{sat}}(M_{\mathrm{h}})&=\frac{a}{2} \times  \left[1 - \mathrm{erf}(\log M_{\mathrm{h}}-b)\right] \\
&+ \frac{c}{2} \times  \left[ 1-\mathrm{erf}(b-\log M_{\mathrm{h}})\right]. \label{equ:p_sat}
\end{aligned}
\end{equation} 
This model reduces the number of satellite ELGs in the massive halos, while retaining satellite ELGs in the low-mass halos. In this way, we keep all the normal central galaxies and randomly select some normal satellite galaxies as ELG candidates according to their probability $P_{\mathrm{sat}}$. At a fixed stellar mass, the satellite fraction of the ELG candidates is identical to the final ELG mock samples we want to get, and the only difference between the two samples is the number density. In the top left panel of Figure \ref{fig:number_boost}, we present the number densities $n^{\mathrm{can}}_{\mathrm{cen}}$ and $n^{\mathrm{can}}_{\mathrm{sat}}$ of ELG candidates as a function of stellar mass. We can notice that the fraction of normal satellite galaxies selected as ELG candidates gradually decreases as the stellar mass increases. 

The second step is to select the ELGs. In this selection, we only need to adjust the number density of galaxies in the simulation to match the ELG SMF in the observations. It is effectively a down-sampling process. We set 45 bins in logarithmic spaces from $10^{7.5}$ to $10^{12}$ $M_{\odot}\, h^{-1}$ with equal widths of $\Delta \log M_{\ast} = 0.1$. In each bin, we calculate the mean number density of ELGs $\bar{n}^{\mathrm{ELG}}_{\mathrm{obs}}\left(M_{\ast}\right)$ in observation (See Paper I for the SMF of ELGs) and ELG candidates $\bar{n}^{\mathrm{ELG}}_{\mathrm{can}}\left(M_{\ast}\right)$ in simulation. Based on the ratio between the two number densities, we can define a probability $F^{\mathrm{ELG}}\left(M_{\ast}\right)$
\begin{equation}
F^{\mathrm{ELG}}\left(M_{\ast}\right) = \frac{\bar{n}^{\mathrm{ELG}}_{\mathrm{obs}}\left(M_{\ast}\right)}{\bar{n}^{\mathrm{ELG}}_{\mathrm{can}}\left(M_{\ast}\right)}. \label{equ:f(M_star)}
\end{equation}   
For each ELG candidate in the simulation, we can assign a probability $F^{\mathrm{ELG}}\left(M_{\ast}\right)$ for it. A random selection is then performed to pick up the final ELG sample in the simulation. The stellar mass distributions of the selected ELGs $n^{\mathrm{ELG}}_{\mathrm{cen}}$ and $n^{\mathrm{ELG}}_{\mathrm{sat}}$ are presented in the top right panel of Figure \ref{fig:number_boost} as solid lines.

Our model has a total of eight parameters $\{\alpha, \beta, M_0, k, \sigma, a, b, c \}$, but there are only seven free parameters as $a$ is set to 1 in Paper I. Using the auto and cross correlation functions of ELGs and LRGs, Paper I simultaneously constrained the SHMR for normal galaxies and the $P_{\mathrm{sat}}$ model. The best-fit parameters of Paper I are listed in the first eight columns of Table \ref{tab:parameters}. 

Taking these parameters, we make a model prediction for the projected correlation functions $\boldsymbol{w}_{\mathrm{p}}$ shown in Figure \ref{fig:wp_noconformity}. We can see that the model is sufficient to reproduce the LRGxELG cross correlation functions. But at $r_{\mathrm{p}}<0.3$ $\mathrm{Mpc}\,h^{-1}$, the observed ELG auto correlation is clearly much higher than the prediction of the model. We note that Paper I has fixed the parameter $a$ to 1, which means no reduction of the ELG satellite fractions in typical halos of $M_{\mathrm{h}} \approx 10^{12}$ $M_{\odot}\, h^{-1}$ that host central ELGs. Namely, the ELG satellite fraction in these halos is the same as that of normal galaxies. Nevertheless, the model is still difficult to match the observed ELG auto correlation on small scales. Therefore, we are considering to incorporate the conformity effect in the original SHAM approach.

\begin{deluxetable*}{cccccccccc}
	\tablenum{2}
	\tablecaption{All parameters in the model. The first eight parameters correspond to the SHMR and $P_{\mathrm{sat}}$ models in Paper I. Except for the parameter $a$ that is fixed to 1, the remaining seven parameters are the best-fit values. The last one is the best-fit conformity parameter $K_{\mathrm{conf}}$ in this work.}
	\label{tab:parameters}
	\tablehead{ \colhead{$\log M_0 \left[M_{\odot}\, h^{-1}\right]$} &
		\colhead{$\alpha$} & \colhead{$\beta$} &\colhead{$\log k$} &\colhead{$\sigma$} &\colhead{$a$}&\colhead{$b$} &\colhead{$c$} &\colhead{$K_{\mathrm{conf}} \; \mathrm{\left(This \, work\right)}$}
	}
	\startdata
	12.07 & 0.37 & 2.61 & 10.36 & 0.21 & 1.00 (fixed) & 12.55 & 0.04 & $5.21^{+0.54}_{-0.50}$ \\
	\enddata
	%\tablecomments{}
\end{deluxetable*}

%\newpage
\subsection{Model conformity effect} \label{sec:AM_conformity}
We aim to propose a concise empirical model of ELG conformity. As argued in the previous subsection, without conformity, we are unable to reproduce the strong auto correlation of ELGs observed on small scales ($r_{\mathrm{p}}<0.3$ $\mathrm{Mpc}\,h^{-1}$), even if the satellite probability $P_{\rm sat}$ is already close to 1 for relevant halos of mass $M_{\mathrm{h}}<10^{12}$ $M_{\odot}\, h^{-1}$. This fact indicates that we need to include some sort of effect, such as galactic conformity, to increase the number of close pairs. As a working hypothesis, we boost the number of satellite ELGs in those halos that host central ELGs (central ELG halos for short in the rest of the paper). 

Firstly, we implement the original SHAM method as described in section \ref{sec:two_step} (see also the top two panels in Figure \ref{fig:number_boost}). In this way, we obtain the ELG central and satellite galaxies in the simulation. For the central ELG halos, we label the number densities of  satellite ELGs and of satellite ELG candidates as $n^{\mathrm{ELG}}_{\mathrm{sat},\mathrm{CE}}$ and $n^{\mathrm{can}}_{\mathrm{sat},\mathrm{CE}}$ respectively, and plot them as green solid and dashed curves in the bottom left panel of Figure \ref{fig:number_boost}. Compared to the total number density $n^{\mathrm{ELG}}_{\mathrm{sat}}$ of satellite ELGs, the fraction of the population in central ELG halos is quite small (less than 3\%), as only a small fraction of central galaxies are qualified for observed ELGs. Comparing the solid and dashed green curves, we find that in the central ELG halos, there are still many satellite ELG candidates that were not selected because only a small fraction is supposed to be included in the observation (cf. Equation \ref{equ:f(M_star)}).

To fit the one-halo term of the ELG auto correlation, we should boost the number density $n^{\mathrm{ELG}}_{\mathrm{sat},\mathrm{CE}}$ to increase the central-satellite and satellite-satellite pair counts of ELGs. To achieve this, we can convert more satellite ELG candidates into final ELGs. Here we introduce $K_{\mathrm{conf}}$ as a free parameter that controls the conversion. At a given stellar mass $M_{\ast}$, we boost the number of the satellite ELG population in the central ELG halos  by a factor of $K_{\mathrm{conf}}$ over the standard satellite occupation number, up to the limit allowed by the remaining satellite ELG candidates. The boosted number density of ELGs $n^{\mathrm{conf}}_{\mathrm{sat}}\left(M_{\ast}\right)$ can be written as  
\begin{equation}
\begin{aligned}
n^{\mathrm{conf}}_{\mathrm{sat},\mathrm{CE}}\left(M_{\ast}\right) 
= \mathrm{min} \{&K_{\mathrm{conf}} \times n^{\mathrm{ELG}}_{\mathrm{sat},\mathrm{CE}}\left(M_{\ast}\right), \\ &\left[n^{\mathrm{can}}_{\mathrm{sat},\mathrm{CE}}\left(M_{\ast}\right) - n^{\mathrm{ELG}}_{\mathrm{sat},\mathrm{CE}}\left(M_{\ast}\right) \right] \}, \label{equ:Nconf}
\end{aligned}
\end{equation}
Since the number of remaining candidates is finite, we take the minimum value between $K_{\mathrm{conf}} \times n^{\mathrm{ELG}}_{\mathrm{sat},\mathrm{CE}}$ and $\left[n^{\mathrm{can}}_{\mathrm{sat},\mathrm{CE}} - n^{\mathrm{ELG}}_{\mathrm{sat},\mathrm{CE}} \right]$ as final $n^{\mathrm{conf}}_{\mathrm{sat},\mathrm{CE}}$. Eventually, we can randomly select galaxies with the number density of  $n^{\mathrm{conf}}_{\mathrm{sat},\mathrm{CE}}\left(M_{\ast}\right)$ from the remaining candidates to augment the satellite ELG population in central ELG halos.

In the bottom right panel of Figure \ref{fig:number_boost}, we set $K_{\mathrm{conf}} = 10$ and illustrate the above process. We observe that the number density of the satellite ELG population in central ELG halos, ($n^{\mathrm{ELG}}_{\mathrm{sat},\mathrm{CE}} + n^{\mathrm{conf}}_{\mathrm{sat},\mathrm{CE}}$), is significantly boosted. In particular, the number density of galaxies has reached the upper limit at stellar mass $\sim 10^{9.5}$ $M_{\odot}$. At low- and high-mass ends, there is still ample room for increasing $n^{\mathrm{conf}}_{\mathrm{sat},\mathrm{CE}}$. Moreover, it should be pointed out that this process can cause the ELG SMF in the simulation to be slightly higher by less than 2 percent. Nevertheless, as we will show in the next Section, the impact of this systematic on the large-scale clustering is also negligible compared to the current measurement accuracy.

\section{Results} \label{sec:Results}
\subsection{Fitting with conformity} \label{sec:Fitting_with_conformity}

\begin{figure}
	\centering
	\includegraphics[scale=0.8]{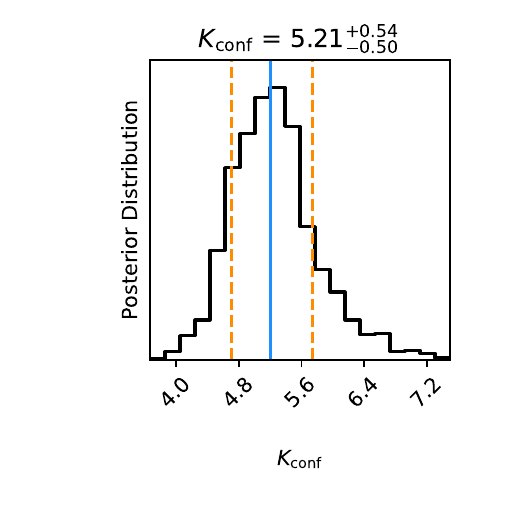}
	\caption{Posterior probability distribution for the conformity model. The median value and its $1\sigma$ confidence interval are also shown.}
	\label{fig:posterior_Kconf}
\end{figure}

\begin{figure*}
	\centering
	\includegraphics[scale=0.6]{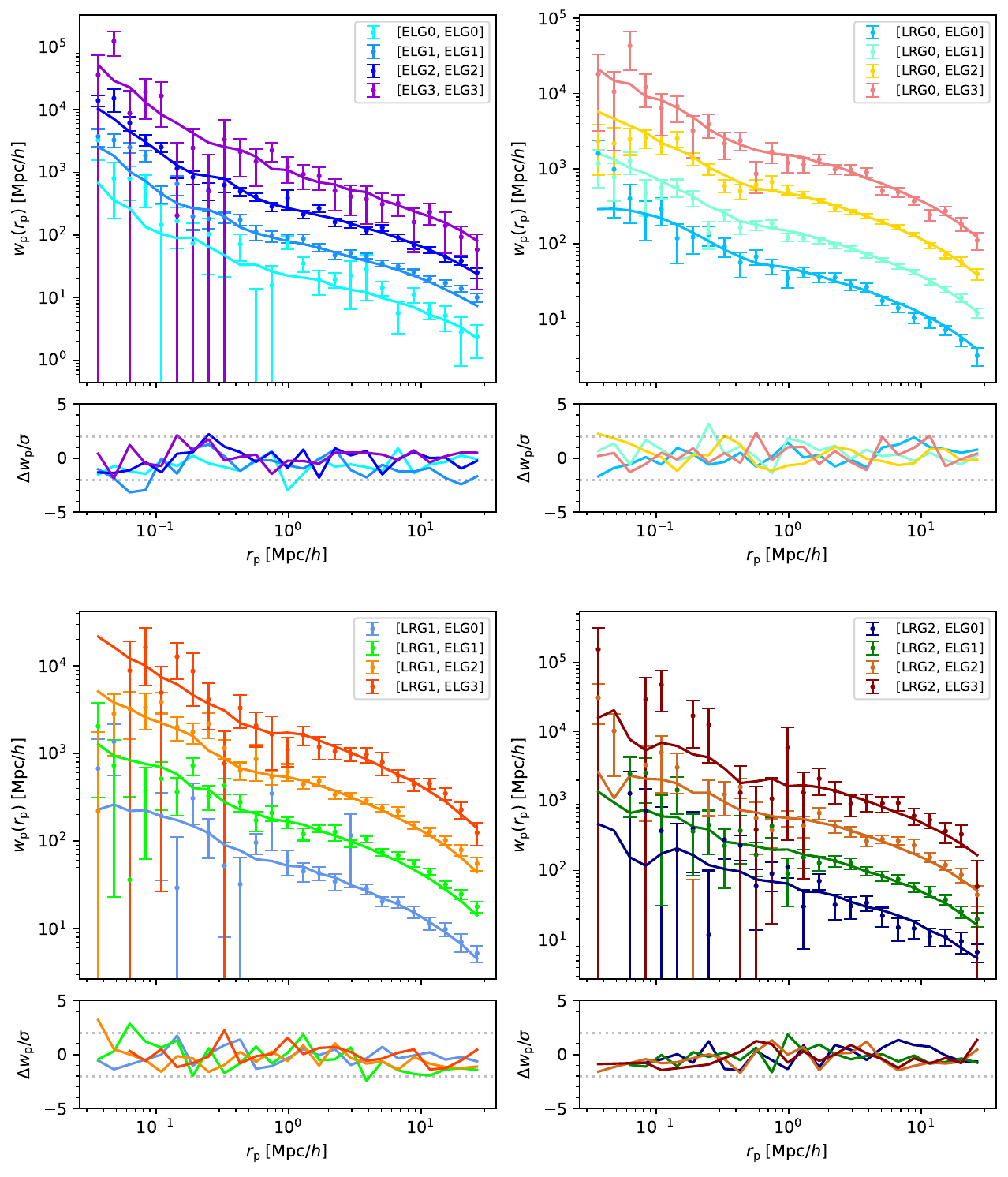}
	\caption{Similar to Figure \ref{fig:wp_noconformity}, but the fitting results for $\boldsymbol{w}_{\mathrm{p}}$ with the conformity model. The solid lines represent the best-fit $\boldsymbol{w}^{\mathrm{mod}}_{\mathrm{p}}$ model.}
	\label{fig:wp_conformity}
\end{figure*}

\begin{figure*}
	\centering
	\includegraphics[scale=0.6]{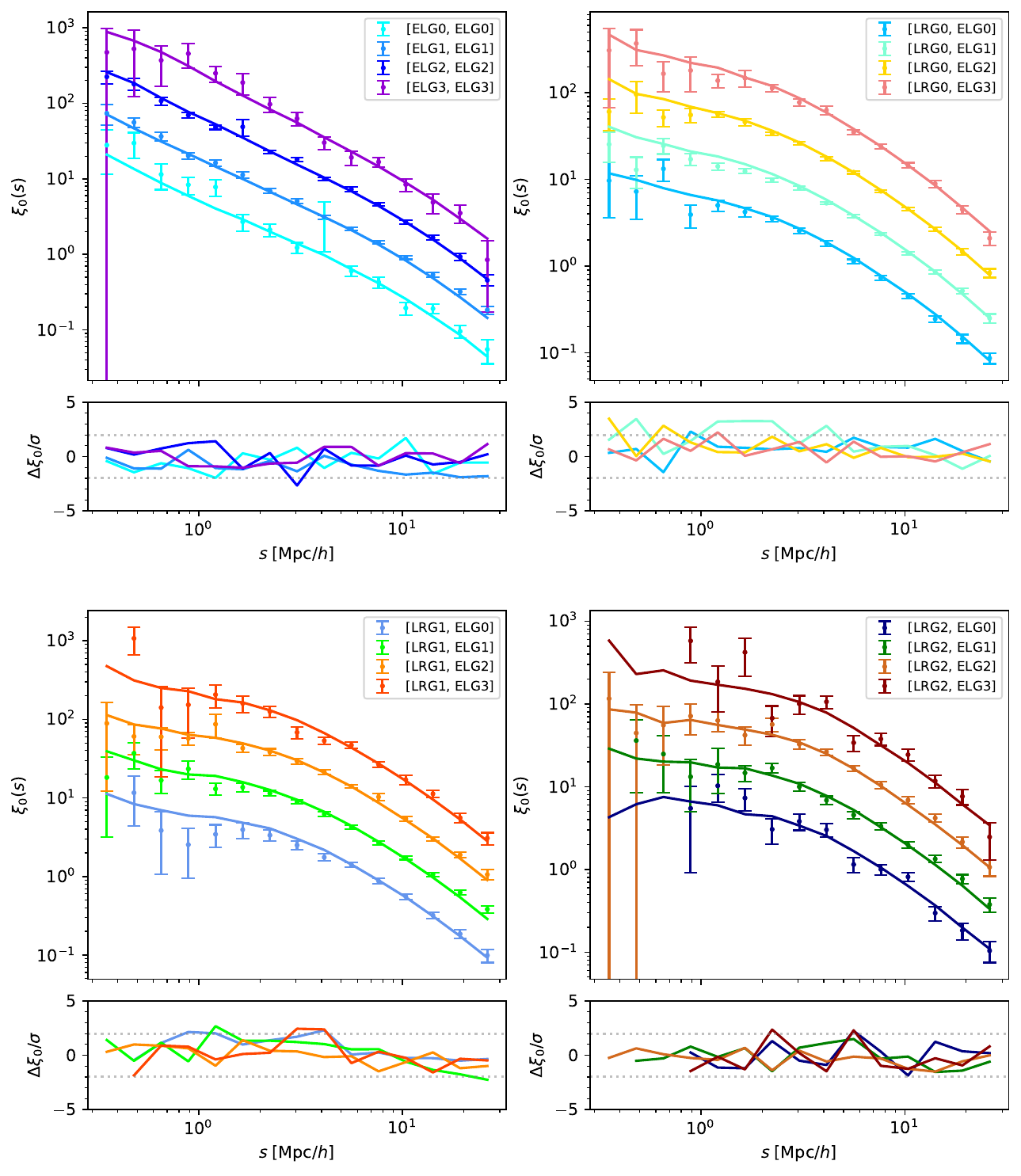}
	\caption{Similar to Figure \ref{fig:wp_conformity}, but the fitting results for monopole $\boldsymbol{\xi}_0$ with the conformity model. The solid lines represent the best-fit $\boldsymbol{\xi}^{\mathrm{mod}}_0$ model. For the purpose of clarity, we multiply each $\boldsymbol{\xi}_0$ by a factor of $3^n$ where $n=0,1,2,3$ correspond to the four ELG subsamples $0,1,2,3$ respectively.}
	\label{fig:xi0_conformity}
\end{figure*}

\begin{figure*}
	\centering
	\includegraphics[scale=0.6]{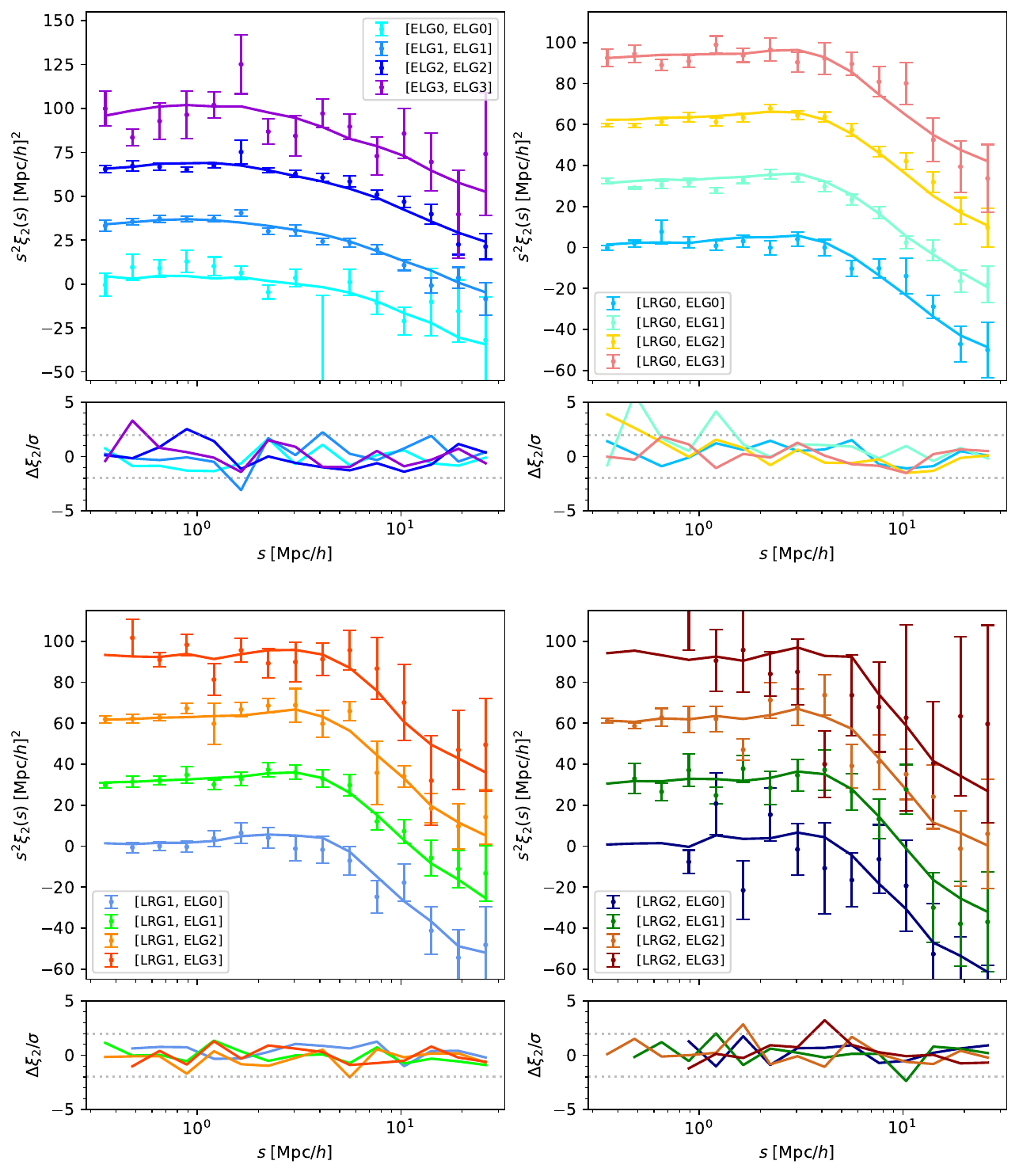}
	\caption{Similar to Figure \ref{fig:xi0_conformity}, but the fitting results for quadrupole $\boldsymbol{\xi}_2$ with the conformity model. For the purpose of clarity, we add each $\boldsymbol{\xi}_2$ by a constant of $30\times n$ where $n=0,1,2,3$ correspond to ELG subsamples $0,1,2,3$ respectively.}
	\label{fig:xi2_conformity}
\end{figure*}

\begin{figure*}
	\centering
	\includegraphics[scale=0.6]{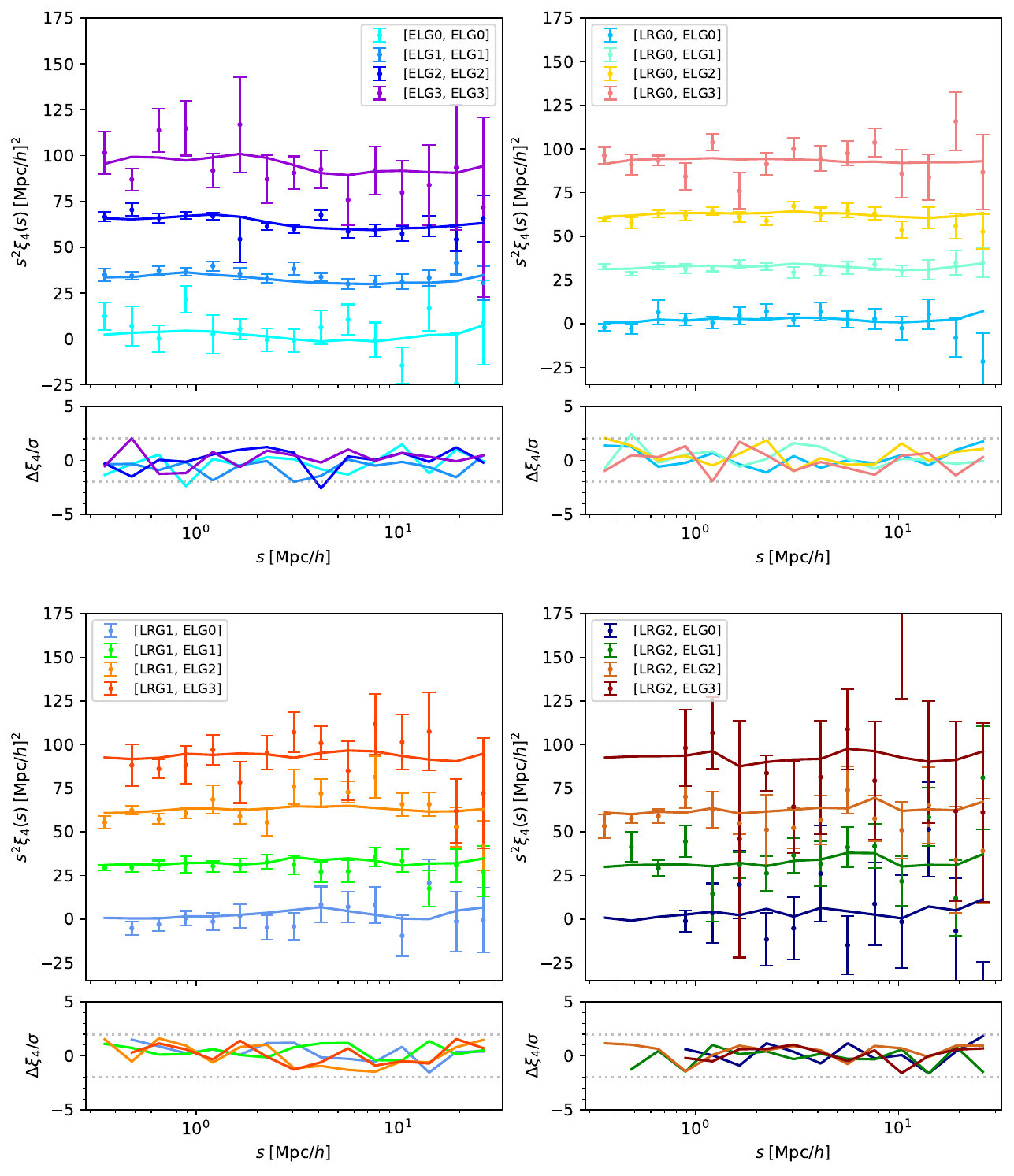}
	\caption{Similar to Figure \ref{fig:xi0_conformity}, but the fitting results for hexadecapole $\boldsymbol{\xi}_4$ with the conformity model. For the purpose of clarity, we add each $\boldsymbol{\xi}_4$ by a constant of $30\times n$ where $n=0,1,2,3$ correspond to ELG subsamples $0,1,2,3$ respectively.}
	\label{fig:xi4_conformity}
\end{figure*}

\begin{figure*}
	\centering
	\includegraphics[scale=0.6]{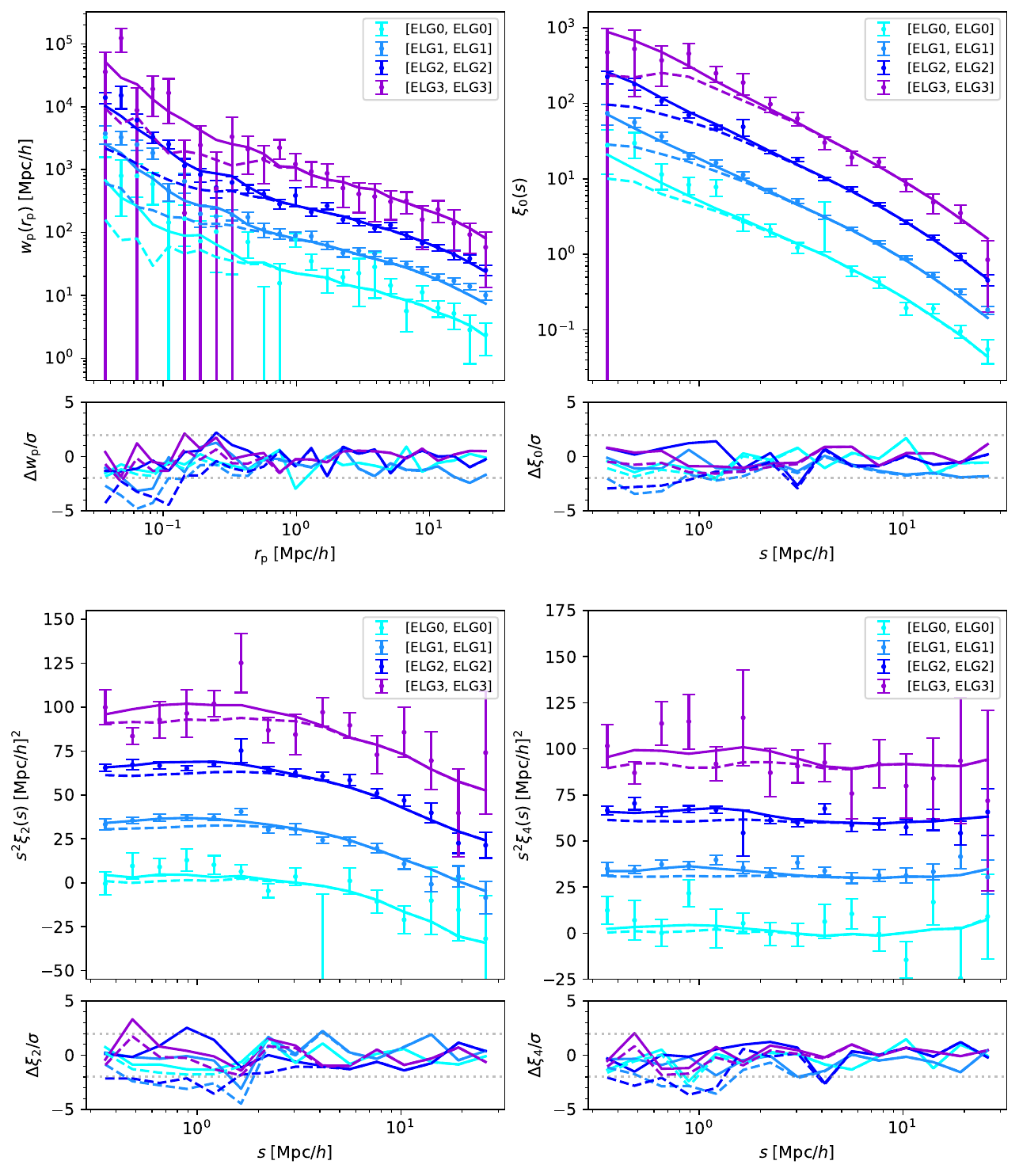}
	\caption{Similar to Figure \ref{fig:wp_noconformity}, \ref{fig:wp_conformity}, \ref{fig:xi0_conformity}, \ref{fig:xi2_conformity}, \ref{fig:xi4_conformity},  auto correlation functions of all ELGs in real and redshift spaces are displayed for side-by-side comparison. The data points with error bars represent the observation. The solid (dashed) lines indicate the model predictions with (without) the conformity effect. With the introduction of ELG conformity, the model improves significantly on small scales.}
	\label{fig:all_auto_correlation}
\end{figure*}

\begin{figure*}
	\centering
	\includegraphics[scale=0.6]{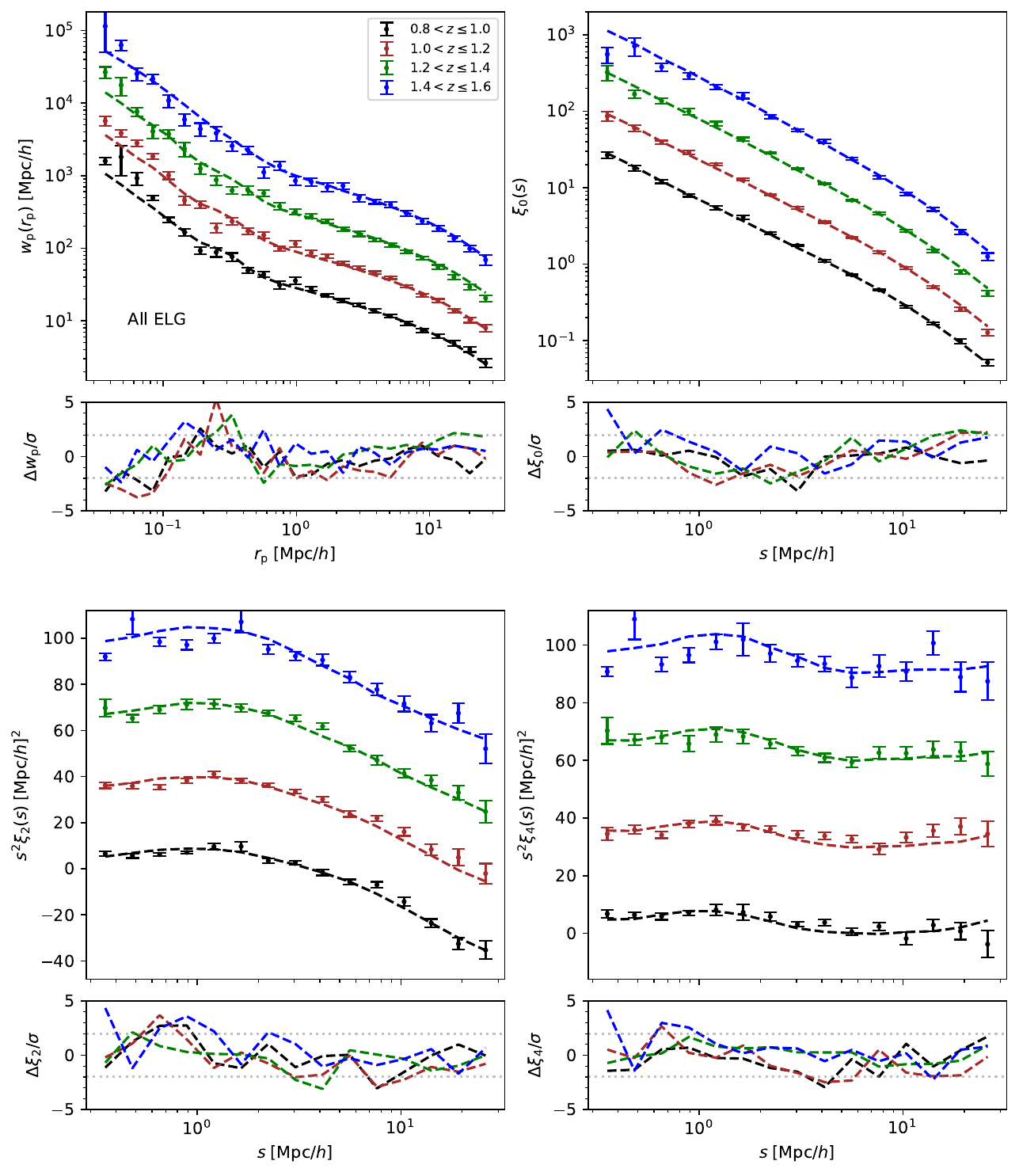}
	\caption{Evolution of the auto correlation functions for all ELGs from $z=0.8$ to $z=1.6$. The four panels correspond to the projected correlation function $\boldsymbol{w}_{\mathrm{p}}$, monopole $\boldsymbol{\xi}_0$, quadrupole $\boldsymbol{\xi}_2$ and hexadecapole $\boldsymbol{\xi}_4$, respectively. The different colored data points indicate the four different redshift intervals. The dashed lines represent the results of model predictions. For the purpose of clarity, each $\boldsymbol{w}_{\mathrm{p}}$ and $\boldsymbol{\xi}_0$ is multiplied by a factor of $3^n$, while each $\boldsymbol{\xi}_2$ and $\boldsymbol{\xi}_4$ is added by a constant of $30\times n$, where $n=0,1,2,3$ correspond to four redshift bins. The differences between model and data, scaled by the measurement errors, are also plotted at the bottom of each panel.}
	\label{fig:evo_wpxi_conformity}
\end{figure*}

\begin{figure*}
	\centering
	\includegraphics[scale=0.8]{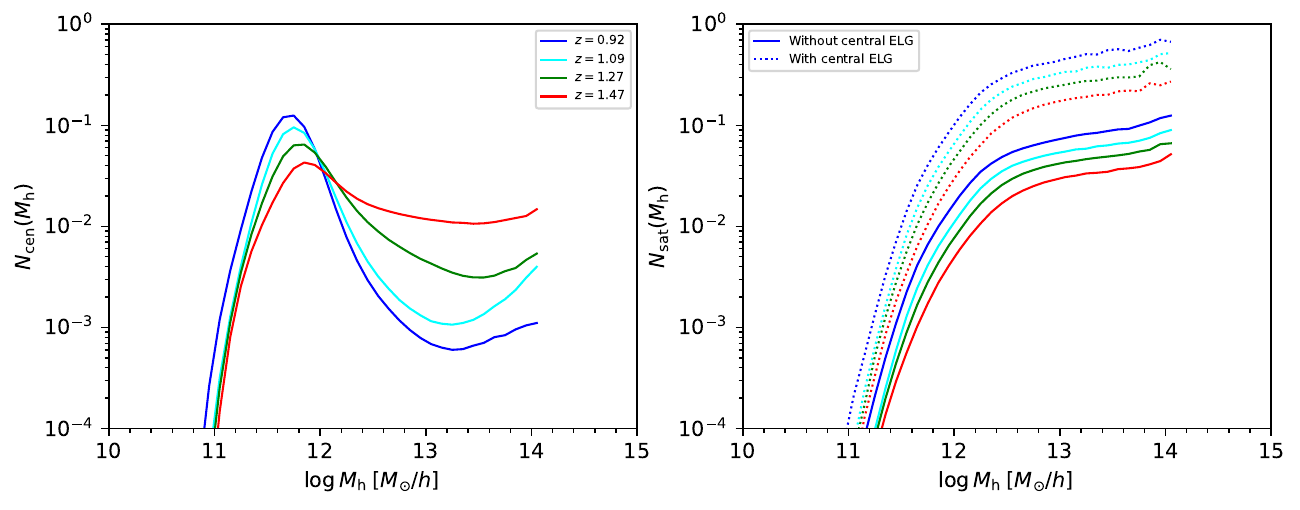}
	\caption{Evolution of the HOD of ELGs in the redshift range $0.8<z<1.6$. Different colors indicate different redshift intervals. The values of the HODs are measured by averaging 1000 random realizations in the simulation. The central occupations $N_{\rm{cen}}(M_{\rm{h}})$ are displayed as solid lines in the left panel. The satellite occupation $N_{\rm{sat}}(M_{\rm{h}})$ shown in the right panel can be divided into two cases. For the halos whose central galaxies are not ELGs, their $N_{\rm{sat}}$ are shown as solid lines. While for the halos with central ELGs, their $N_{\rm{sat}}$ are presented as dotted lines. The difference in $N_{\rm{sat}}$ between the two cases reflects the effect of the conformity model.}
	\label{fig:conformity_HOD_simulation}
\end{figure*}

In this Section, we aim to determine the free parameter $K_{\mathrm{conf}}$. In real space, 4 ELG projected auto correlations $\boldsymbol{w}^{\mathrm{obs}}_{\mathrm{p},\mathrm{ELG}}$ and 12 LRGxELG projected cross correlations $\boldsymbol{w}^{\mathrm{obs}}_{\mathrm{p},\mathrm{LRGxELG}}$ are used in the fitting process. The resulting $\chi^2_{w_{\mathrm{p}},xy}$ can be expressed as 
\begin{equation}
\begin{aligned}
\chi^2_{w_{\mathrm{p}},xy}&=\\
\sum_{k=1}^{N_{r_{\mathrm{p}}}}&\sum_{m=1}^{N_{r_{\mathrm{p}}}}\left(w^{\mathrm{obs}}_{\mathrm{p},xy,k}-w^{\mathrm{mod}}_{\mathrm{p},xy,k}\right)C^{-1}_{xy,km}\left(w^{\mathrm{obs}}_{\mathrm{p},xy,m}-w^{\mathrm{mod}}_{\mathrm{p},xy,m}\right), \label{equ:chi2_wp}
\end{aligned}
\end{equation}
where $x$ and $y$ indicate different subsamples ($x=y$ for auto correlation), $w^{\mathrm{obs}}_{\mathrm{p}}$ and $w^{\mathrm{mod}}_{\mathrm{p}}$ represent the observed and modeled correlation functions. 

Similarly, we also include the clustering in redshift space in the fit. For subsample $x$ and $y$, the corresponding $\chi^2_{\xi_{l},xy}$ is

\begin{equation}
\chi^2_{\xi_{l},xy}=\sum_{k=1}^{N_{s}}\sum_{m=1}^{N_{s}}\left(\xi^{\mathrm{obs}}_{l,xy,k}-\xi^{\mathrm{mod}}_{l,xy,k}\right)C^{-1}_{xy,km}\left(\xi^{\mathrm{obs}}_{l,xy,m}-\xi^{\mathrm{mod}}_{l,xy,m}\right), \label{equ:chi2_xi}
\end{equation}
where $\xi_{l}$ denote the $l$-th multiple moment.

In this way, the total $\chi^2$ for all combinations of subsamples can be written as
\begin{equation}
\begin{aligned}
\chi^2 =
 &\sum_{x}\chi^2_{w_{\mathrm{p}},xx} + \sum_{x}\sum_{y}\chi^2_{w_{\mathrm{p}},xy} \\ 
 &+ \sum_{x} \sum_{l}\chi^2_{\xi_{l},xx} + \sum_{x}\sum_{y} \sum_{l}\chi^2_{\xi_{l},xy} , \label{equ:chi2_all}
\end{aligned}
\end{equation}
where $x \in $ [ELG0, ELG1, ELG2, ELG3], $y \in $ [LRG0, LRG1, LRG2] and $l \in $ [0, 2, 4]. We perform a Markov Chain Monte Carlo (MCMC) analysis utilizing {\tt\string emcee} \citep{2013PASP..125..306F}. The posterior distribution of $K_{\mathrm{conf}}$ in the parameter space is displayed in Figure \ref{fig:posterior_Kconf} and the last column of Table \ref{tab:parameters}.

The fitting of all these correlation functions yields a reduced $\chi^2 = 1184.07/1076 = 1.10$. In Figure \ref{fig:wp_conformity}, the best-fit $\boldsymbol{w}^{\mathrm{mod}}_{\mathrm{p}}$ models are plotted as solid lines. Compared to the top left panel of Figure \ref{fig:wp_noconformity}, the model for ELG auto correlations at $r_{\mathrm{p}}<0.5$ $\mathrm{Mpc}\,h^{-1}$ is greatly improved. To achieve such strong clustering, the number of satellite ELGs with respect to an central ELG should be increased by a factor of $5.21$. In the case of LRGxELG cross correlations, the conformity effect has a minimal impact on the results as expected. This is because the majority of LRGs are massive central galaxies, while the conformity effect primarily enhances the number of satellite ELGs in low-mass halos that host central ELGs. As a result, this effect has little influence on the LRGxELG cross correlations.

In addition to $\boldsymbol{w}^{\mathrm{mod}}_{\mathrm{p}}$, we also check the best-fit correlation functions $\boldsymbol{\xi}_0$, $\boldsymbol{\xi}_2$ and $\boldsymbol{\xi}_4$ in redshift space, as displayed in Figure \ref{fig:xi0_conformity}, \ref{fig:xi2_conformity} and \ref{fig:xi4_conformity}, respectively. We also compare our model predictions with and without galactic conformity with the auto correlations of {\it all} ELGs both in real space and in redshift space (Figure \ref{fig:all_auto_correlation}). With the galactic conformity, we are now able to get a good fitting result down to $s\sim 0.3$ $\mathrm{Mpc}\,h^{-1}$. Most of the deviations between the data points and the models fall within the $2 \sigma$ range. The results indicate that our model agrees well with all data points from large to the smallest scales ($r_{\mathrm{p}} \sim 0.03$ $\mathrm{Mpc}\,h^{-1}$ in real space and $s \sim 0.3$ $\mathrm{Mpc}\,h^{-1}$ in redshift space) where the correlations can be reasonably measured. 

Recently, \cite{abacusELG_Rocher} also carefully analyzed the conformity bias of the ELGs in the One-Percent Survey. They extend the HOD framework to include a conformity model in which satellite ELGs are only allowed to exist in the central ELG halos. Although the ELG clustering can be well described by their model, the introduction of conformity changes the original HOD parameters. Especially, the satellite fraction is reduced from $\sim 12 \%$ to $\sim 2 \%$. By the construction, the cross correlation between ELGs and LRGs would be expected to be very low (even negative) on scales smaller than the typical LRG halo size ($\sim 0.4$ $\mathrm{Mpc}\,h^{-1}$), which is contradictory with the cross correlation measurement (Figure \ref{fig:wp_conformity}).  

%\newpage
\subsection{Check with all ELG samples} \label{sec:all_ELGs}

In practice, for cosmological analysis, one usually measures the clustering of the full ELG sample. Therefore, we need to verify that our model is valid for the entire ELG sample, but not just for the subsamples binned by stellar mass. In Figure \ref{fig:evo_wpxi_conformity}, we present the evolution of the ELG auto correlations from redshift $z=0.8$ to $z=1.6$. The ELG in each redshift interval is the full sample without stellar mass binning. The correlation functions in both real space and redshift space are shown together in Figure \ref{fig:evo_wpxi_conformity}. 

The model predictions are plotted as the dashed lines. In general, the model predictions are in very good agreement with the data. The only difference appears to be in the $\boldsymbol{w}_{\mathrm{p}}$ at $r_{\mathrm{p}} \sim 0.4$ $\mathrm{Mpc}\,h^{-1}$, where the one observed data point at each redshift is always below the model prediction. We notice that the drop scale moves to a larger scale as the redshift increases. If there is some angular observational effect that is not fully corrected in the correlation analysis on scale $0.45$ $\mathrm {arcmin}$, we would expect the scale of the drop to shift with the redshift. Interestingly, this scale is nearly the same as the mean separation of the fibers, and therefore we suspect that the drop may be due to uncorrected fiber collisions, even though the PIP and ANG up-weighting have been taken into account in our measurement. Apart from this specific difference, our model is in good agreement with the observational data within $2\sigma$. This means that the model for both the ELG-halo connection and the conformity has a weak dependence on redshift. With the parameters obtained at $z\sim 0.9$, we can extend the model to $z \sim 1.5$ without requiring further modifications to the model parameters.

\subsection{ELG HODs with conformity} \label{sec:HOD_conformity}
We further investigate the HOD of ELGs after introducing the conformity model. We apply the new SHAM method to the simulation and perform 1000 random realizations. The values of HOD presented in Figure \ref{fig:conformity_HOD_simulation} are the average of these realizations. We show the central occupations $N_{\rm{cen}}(M_{\rm{h}})$ as solid lines in the left panel. We can see that the $N_{\rm{cen}}$ peaks around $10^{12}$ $M_{\odot}\, h^{-1}$ and decreases towards the low- and high-mass ends. The position of the peak also gradually increases with redshift. The shape and evolution of the $N_{\rm{cen}}$ has been analyzed in detail in Paper I. By our design, the conformity model does not have any effect on the $N_{\rm{cen}}$.

As for the satellite occupations $N_{\rm{sat}}(M_{\rm{h}})$, we discuss them in two cases. First, for the halos whose central galaxies are not ELGs, we display their $N_{\rm{sat}}$ as solid lines in the right panel of Figure \ref{fig:conformity_HOD_simulation}. In this case, the $N_{\rm{sat}}$ is actually equivalent to the original model without conformity effect. Next, we also present the $N_{\rm{sat}}$ for the central ELG halos as the dotted lines. We can notice that although the conformity model increases the number of satellite ELGs in the central ELG halos by a factor of $\sim 5$, the $N_{\rm{sat}}$ is still smaller than 1, even in the massive halos with mass $\sim 10^{14}$ $M_{\odot}\, h^{-1}$. In other words, the number of satellite ELGs is always small ($<1$), even in the central ELG halos. On the one hand, the $P_{\rm{sat}}$ model is only $\sim 0.04$ at the massive end, and thus the number of ELG candidates in the large halo is small. On the other hand, the observed ELG SMF limits the number of ELGs that can be ultimately selected. As the result, this conformity effect does not significantly increase the overall number of satellite ELGs in the model. For example, at $z \sim 0.9$, the ELG satellite fraction in the model has only changed from 15.7\% to 17.6\% when conformity is included.

In our model, the best-fitting conformity parameter $K_{\mathrm{conf}}$ is $5.21^{+0.54}_{-0.50}$, which appears to be a large effect compared to the previous results about galactic conformity \citep[e.g.,][]{2006MNRAS.366....2W}. As the HODs in Figure \ref{fig:conformity_HOD_simulation} show, DESI ELGs represent less than 10\% of the galaxies at a given stellar mass, and they are the bluest galaxies with the strongest [O\,II] emission lines. However, previous studies usually divided galaxies into two populations, red and blue, according to their colors. This may result in a larger population of blue (or star-forming) galaxies than the DESI ELG samples. The measurements from \cite{2022ApJ...928...10G} also demonstrate that an ELG sample with the higher [O\,II] luminosity presents a stronger conformity signal than the one with the lower [O\,II] luminosity. This explains why the DESI ELGs that are prominent [O\,II] emitters show the pronounced conformity effect. 

%\newpage
\section{Conclusion} \label{sec:Conclusion}
In this study, we investigate the conformity effect of ELGs samples from the DESI One-Percent survey and improve the SHAM method developed in Paper I for constructing the ELG-halo connection. The main conclusions of this paper can be summarized as follows.

\begin{enumerate}
\item We have shown that the strong clustering of ELGs on small scales both in real ($< 0.3\mathrm{Mpc}\,h^{-1}$) and redshift ($s< 1 \mathrm{Mpc}\,h^{-1}$) spaces can not be explained by the abundance matching model developed in Paper I.  Although there are sufficient number of subhalos that are massive enough to accommodate ELGs in halos of mass about $10^{12}$$M_{\odot}\, h^{-1}$ (i.e. ELG candidates in Section \ref{sec:AM}), these candidates are not selected because of the uniform down-sampling in selecting ELGs to satisfying the observed number density (Equation \ref{equ:f(M_star)}). 
	
	\item We refine our model by incorporating the galactic conformity effect. For the satellite ELGs in the central ELG halos, we increase their number density by a factor of $K_{\mathrm{conf}}$ or up to the maximum limits allowed by the number density of the remaining satellite ELG candidates. 
	
	\item By simultaneously fitting the ELG auto and LRGxELG cross correlations in both real and redshift space, we show that the best-fit value of $K_{\mathrm{conf}}=5.21^{+0.54}_{-0.50}$ is well determined. Using this simple conformity model, we can significantly improve the model predictions of $\boldsymbol{w}_{\mathrm{p}}$ in real space and the multipole moments in redshift space on small scales.
	
	\item We also use this model to predict the clustering of the whole ELG population at different redshifts from $z=0.8$ to $z=1.6$. Although our model is derived from the subsamples binned with stellar mass at $0.8<z<1.0$, the prediction of our model agrees well with the observed clustering at $r_{\mathrm{p}} > 0.03$ $\mathrm{Mpc}\,h^{-1}$ in real space and $s > 0.3$ $\mathrm{Mpc}\,h^{-1}$ in redshift space, of the whole ELG population over the entire redshift range $0.8<z<1.6$ explored by DESI.
	
	\item We further analyzed the effect of conformity model on the HODs of ELGs. We find that although the number of satellite ELGs is boosted by a factor of $\sim 5$ in the central ELG halos, the satellite occupation number is still smaller than 1. Compared to the normal galaxies, the number of satellite ELGs is very small in each halo, even in the halos subject to the conformity effect. As the number of the central ELG halos is small, the galactic conformity increases ELGs only by a small fraction in the total population or even in the satellite population. This effect does boost the clustering on small scales, but does not alter the correlation on large scales or the observed SMF.
\end{enumerate} 

By combining the SHAM model with the ELG conformity model, we can effectively recover the correlation functions of ELGs given the current uncertainties. We believe that this model can be conveniently applied to create high-quality ELG lightcone mock catalogs for DESI and other surveys mainly based on ELGs.

The galactic conformity is expected to be a continuous function of the physical properties. We plan to use future DESI redshift samples, which will be much larger than the one used here, to explore the galactic conformity in the space of physical properties. High order clustering will be used to increase the observational power to discriminate among different assumptions on the conformity. In the theory, we will incorporate hydrodynamical simulations to adopt more physical assumptions on the conformity. With these efforts, we expect to elaborate the galactic conformity model in the near future.  

\section*{Acknowledgments}
We thank John Peacock and Jeremy Tinker for their great help during the DESI Collaboration Wide Review. 

The work is supported by NSFC (12133006, 11890691, 11621303), by grant No. CMS-CSST-2021-A03, and by 111 project No. B20019. We gratefully acknowledge the support of the Key Laboratory for Particle Physics, Astrophysics and Cosmology, Ministry of Education. This work made use of the Gravity Supercomputer at the Department of Astronomy, Shanghai Jiao Tong University.

This material is based upon work supported by the U.S. Department of Energy (DOE), Office of Science, Office of High-Energy Physics, under Contract No. DE–AC02–05CH11231, and by the National Energy Research Scientific Computing Center, a DOE Office of Science User Facility under the same contract. Additional support for DESI was provided by the U.S. National Science Foundation (NSF), Division of Astronomical Sciences under Contract No. AST-0950945 to the NSF’s National Optical-Infrared Astronomy Research Laboratory; the Science and Technology Facilities Council of the United Kingdom; the Gordon and Betty Moore Foundation; the Heising-Simons Foundation; the French Alternative Energies and Atomic Energy Commission (CEA); the National Council of Science and Technology of Mexico (CONACYT); the Ministry of Science and Innovation of Spain (MICINN), and by the DESI Member Institutions: \url{https://www.desi.lbl.gov/collaborating-institutions}. 

The DESI Legacy Imaging Surveys consist of three individual and complementary projects: the Dark Energy Camera Legacy Survey (DECaLS), the Beijing-Arizona Sky Survey (BASS), and the Mayall z-band Legacy Survey (MzLS). DECaLS, BASS and MzLS together include data obtained, respectively, at the Blanco telescope, Cerro Tololo Inter-American Observatory, NSF’s NOIRLab; the Bok telescope, Steward Observatory, University of Arizona; and the Mayall telescope, Kitt Peak National Observatory, NOIRLab. NOIRLab is operated by the Association of Universities for Research in Astronomy (AURA) under a cooperative agreement with the National Science Foundation. Pipeline processing and analyses of the data were supported by NOIRLab and the Lawrence Berkeley National Laboratory. Legacy Surveys also uses data products from the Near-Earth Object Wide-field Infrared Survey Explorer (NEOWISE), a project of the Jet Propulsion Laboratory/California Institute of Technology, funded by the National Aeronautics and Space Administration. Legacy Surveys was supported by: the Director, Office of Science, Office of High Energy Physics of the U.S. Department of Energy; the National Energy Research Scientific Computing Center, a DOE Office of Science User Facility; the U.S. National Science Foundation, Division of Astronomical Sciences; the National Astronomical Observatories of China, the Chinese Academy of Sciences and the Chinese National Natural Science Foundation. LBNL is managed by the Regents of the University of California under contract to the U.S. Department of Energy. The complete acknowledgments can be found at \url{https://www.legacysurvey.org/}. 

Any opinions, findings, and conclusions or recommendations expressed in this material are those of the author(s) and do not necessarily reflect the views of the U. S. National Science Foundation, the U. S. Department of Energy, or any of the listed funding agencies.

The authors are honored to be permitted to conduct scientific research on Iolkam Du'ag (Kitt Peak), a mountain with particular significance to the Tohono O'odham Nation.

\software{Numpy \citep{5725236}, Scipy \citep{4160250}, Matplotlib \citep{4160265}, Astropy \citep{2013A&A...558A..33A,2018AJ....156..123A,2022ApJ...935..167A}, emcee \citep{2013PASP..125..306F}, corner \citep{2016JOSS....1...24F}, Corrfunc \citep{2020MNRAS.491.3022S}}

\section*{Data Availability}
Data points of each figure are available in \url{https://doi.org/10.5281/zenodo.8296757}

\appendix

\bibliography{sv3_elg_conformity}{}

\begin{thebibliography}{}
\expandafter\ifx\csname natexlab\endcsname\relax\def\natexlab#1{#1}\fi
\providecommand{\url}[1]{\href{#1}{#1}}
\providecommand{\dodoi}[1]{doi:~\href{http://doi.org/#1}{\nolinkurl{#1}}}
\providecommand{\doeprint}[1]{\href{http://ascl.net/#1}{\nolinkurl{http://ascl.net/#1}}}
\providecommand{\doarXiv}[1]{\href{https://arxiv.org/abs/#1}{\nolinkurl{https://arxiv.org/abs/#1}}}

\bibitem[{{Alam} {et~al.}(2020){Alam}, {Peacock}, {Kraljic}, {Ross}, \&
  {Comparat}}]{2020MNRAS.497..581A}
{Alam}, S., {Peacock}, J.~A., {Kraljic}, K., {Ross}, A.~J., \& {Comparat}, J.
  2020, \mnras, 497, 581, \dodoi{10.1093/mnras/staa1956}

\bibitem[{{Alexander} {et~al.}(2023){Alexander}, {Davis}, {Chaussidon},
  {Fawcett}, {X. Gonzalez-Morales}, {Lan}, {Y{\`e}che}, {Ahlen}, {Aguilar},
  {Armengaud}, {Bailey}, {Brooks}, {Cai}, {Canning}, {Carr}, {Chabanier},
  {Cousinou}, {Dawson}, {de la Macorra}, {Dey}, {Dey}, {Dhungana}, {Edge},
  {Eftekharzadeh}, {Fanning}, {Farr}, {Font-Ribera}, {Garcia-Bellido},
  {Garrison}, {Gazta{\~n}aga}, {A Gontcho}, {Gordon}, {Medellin Gonzalez},
  {Guy}, {Herrera-Alcantar}, {Jiang}, {Juneau}, {Kara{\c{c}}ayl{\i}}, {Kehoe},
  {Kisner}, {Kov{\'a}cs}, {Landriau}, {Levi}, {Magneville}, {Martini},
  {Meisner}, {Mezcua}, {Miquel}, {Camacho}, {Moustakas},
  {Mu{\~n}oz-Guti{\'e}rrez}, {Myers}, {Nadathur}, {Napolitano}, {Nie},
  {Palanque-Delabrouille}, {Pan}, {Percival}, {P{\'e}rez-R{\`a}fols},
  {Poppett}, {Prada}, {Ram{\'\i}rez-P{\'e}rez}, {Ravoux}, {Rosario},
  {Schubnell}, {Tarl{\'e}}, {Walther}, {Weiner}, {Youles}, {Zhou}, {Zou}, \&
  {Zou}}]{2023AJ....165..124A}
{Alexander}, D.~M., {Davis}, T.~M., {Chaussidon}, E., {et~al.} 2023, \aj, 165,
  124, \dodoi{10.3847/1538-3881/acacfc}

\bibitem[{{Allende Prieto} {et~al.}(2020){Allende Prieto}, {Cooper}, {Dey},
  {G{\"a}nsicke}, {Koposov}, {Li}, {Manser}, {Nidever}, {Rockosi}, {Wang},
  {Aguado}, {Blum}, {Brooks}, {Eisenstein}, {Duan}, {Eftekharzadeh},
  {Gazta{\~n}aga}, {Kehoe}, {Landriau}, {Lee}, {Levi}, {Meisner}, {Myers},
  {Najita}, {Olsen}, {Palanque-Delabrouille}, {Poppett}, {Prada}, {Schlegel},
  {Schubnell}, {Tarl{\'e}}, {Valluri}, {Wechsler}, \&
  {Y{\`e}che}}]{2020RNAAS...4..188A}
{Allende Prieto}, C., {Cooper}, A.~P., {Dey}, A., {et~al.} 2020, Research Notes
  of the American Astronomical Society, 4, 188,
  \dodoi{10.3847/2515-5172/abc1dc}

\bibitem[{{Astropy Collaboration} {et~al.}(2013){Astropy Collaboration},
  {Robitaille}, {Tollerud}, {Greenfield}, {Droettboom}, {Bray}, {Aldcroft},
  {Davis}, {Ginsburg}, {Price-Whelan}, {Kerzendorf}, {Conley}, {Crighton},
  {Barbary}, {Muna}, {Ferguson}, {Grollier}, {Parikh}, {Nair}, {Unther},
  {Deil}, {Woillez}, {Conseil}, {Kramer}, {Turner}, {Singer}, {Fox}, {Weaver},
  {Zabalza}, {Edwards}, {Azalee Bostroem}, {Burke}, {Casey}, {Crawford},
  {Dencheva}, {Ely}, {Jenness}, {Labrie}, {Lim}, {Pierfederici}, {Pontzen},
  {Ptak}, {Refsdal}, {Servillat}, \& {Streicher}}]{2013A&A...558A..33A}
{Astropy Collaboration}, {Robitaille}, T.~P., {Tollerud}, E.~J., {et~al.} 2013,
  \aap, 558, A33, \dodoi{10.1051/0004-6361/201322068}

\bibitem[{{Astropy Collaboration} {et~al.}(2018){Astropy Collaboration},
  {Price-Whelan}, {Sip{\H{o}}cz}, {G{\"u}nther}, {Lim}, {Crawford}, {Conseil},
  {Shupe}, {Craig}, {Dencheva}, {Ginsburg}, {VanderPlas}, {Bradley},
  {P{\'e}rez-Su{\'a}rez}, {de Val-Borro}, {Aldcroft}, {Cruz}, {Robitaille},
  {Tollerud}, {Ardelean}, {Babej}, {Bach}, {Bachetti}, {Bakanov}, {Bamford},
  {Barentsen}, {Barmby}, {Baumbach}, {Berry}, {Biscani}, {Boquien}, {Bostroem},
  {Bouma}, {Brammer}, {Bray}, {Breytenbach}, {Buddelmeijer}, {Burke},
  {Calderone}, {Cano Rodr{\'\i}guez}, {Cara}, {Cardoso}, {Cheedella}, {Copin},
  {Corrales}, {Crichton}, {D'Avella}, {Deil}, {Depagne}, {Dietrich}, {Donath},
  {Droettboom}, {Earl}, {Erben}, {Fabbro}, {Ferreira}, {Finethy}, {Fox},
  {Garrison}, {Gibbons}, {Goldstein}, {Gommers}, {Greco}, {Greenfield},
  {Groener}, {Grollier}, {Hagen}, {Hirst}, {Homeier}, {Horton}, {Hosseinzadeh},
  {Hu}, {Hunkeler}, {Ivezi{\'c}}, {Jain}, {Jenness}, {Kanarek}, {Kendrew},
  {Kern}, {Kerzendorf}, {Khvalko}, {King}, {Kirkby}, {Kulkarni}, {Kumar},
  {Lee}, {Lenz}, {Littlefair}, {Ma}, {Macleod}, {Mastropietro}, {McCully},
  {Montagnac}, {Morris}, {Mueller}, {Mumford}, {Muna}, {Murphy}, {Nelson},
  {Nguyen}, {Ninan}, {N{\"o}the}, {Ogaz}, {Oh}, {Parejko}, {Parley}, {Pascual},
  {Patil}, {Patil}, {Plunkett}, {Prochaska}, {Rastogi}, {Reddy Janga},
  {Sabater}, {Sakurikar}, {Seifert}, {Sherbert}, {Sherwood-Taylor}, {Shih},
  {Sick}, {Silbiger}, {Singanamalla}, {Singer}, {Sladen}, {Sooley},
  {Sornarajah}, {Streicher}, {Teuben}, {Thomas}, {Tremblay}, {Turner},
  {Terr{\'o}n}, {van Kerkwijk}, {de la Vega}, {Watkins}, {Weaver}, {Whitmore},
  {Woillez}, {Zabalza}, \& {Astropy Contributors}}]{2018AJ....156..123A}
{Astropy Collaboration}, {Price-Whelan}, A.~M., {Sip{\H{o}}cz}, B.~M., {et~al.}
  2018, \aj, 156, 123, \dodoi{10.3847/1538-3881/aabc4f}

\bibitem[{{Astropy Collaboration} {et~al.}(2022){Astropy Collaboration},
  {Price-Whelan}, {Lim}, {Earl}, {Starkman}, {Bradley}, {Shupe}, {Patil},
  {Corrales}, {Brasseur}, {N{\"o}the}, {Donath}, {Tollerud}, {Morris},
  {Ginsburg}, {Vaher}, {Weaver}, {Tocknell}, {Jamieson}, {van Kerkwijk},
  {Robitaille}, {Merry}, {Bachetti}, {G{\"u}nther}, {Aldcroft},
  {Alvarado-Montes}, {Archibald}, {B{\'o}di}, {Bapat}, {Barentsen},
  {Baz{\'a}n}, {Biswas}, {Boquien}, {Burke}, {Cara}, {Cara}, {Conroy},
  {Conseil}, {Craig}, {Cross}, {Cruz}, {D'Eugenio}, {Dencheva}, {Devillepoix},
  {Dietrich}, {Eigenbrot}, {Erben}, {Ferreira}, {Foreman-Mackey}, {Fox},
  {Freij}, {Garg}, {Geda}, {Glattly}, {Gondhalekar}, {Gordon}, {Grant},
  {Greenfield}, {Groener}, {Guest}, {Gurovich}, {Handberg}, {Hart},
  {Hatfield-Dodds}, {Homeier}, {Hosseinzadeh}, {Jenness}, {Jones}, {Joseph},
  {Kalmbach}, {Karamehmetoglu}, {Ka{\l}uszy{\'n}ski}, {Kelley}, {Kern},
  {Kerzendorf}, {Koch}, {Kulumani}, {Lee}, {Ly}, {Ma}, {MacBride}, {Maljaars},
  {Muna}, {Murphy}, {Norman}, {O'Steen}, {Oman}, {Pacifici}, {Pascual},
  {Pascual-Granado}, {Patil}, {Perren}, {Pickering}, {Rastogi}, {Roulston},
  {Ryan}, {Rykoff}, {Sabater}, {Sakurikar}, {Salgado}, {Sanghi}, {Saunders},
  {Savchenko}, {Schwardt}, {Seifert-Eckert}, {Shih}, {Jain}, {Shukla}, {Sick},
  {Simpson}, {Singanamalla}, {Singer}, {Singhal}, {Sinha}, {Sip{\H{o}}cz},
  {Spitler}, {Stansby}, {Streicher}, {{\v{S}}umak}, {Swinbank}, {Taranu},
  {Tewary}, {Tremblay}, {de Val-Borro}, {Van Kooten}, {Vasovi{\'c}}, {Verma},
  {de Miranda Cardoso}, {Williams}, {Wilson}, {Winkel}, {Wood-Vasey}, {Xue},
  {Yoachim}, {Zhang}, {Zonca}, \& {Astropy Project
  Contributors}}]{2022ApJ...935..167A}
{Astropy Collaboration}, {Price-Whelan}, A.~M., {Lim}, P.~L., {et~al.} 2022,
  \apj, 935, 167, \dodoi{10.3847/1538-4357/ac7c74}

\bibitem[{{Avila} {et~al.}(2020){Avila}, {Gonzalez-Perez}, {Mohammad}, {de
  Mattia}, {Zhao}, {Raichoor}, {Tamone}, {Alam}, {Bautista}, {Bianchi},
  {Burtin}, {Chapman}, {Chuang}, {Comparat}, {Dawson}, {Divers}, {du Mas des
  Bourboux}, {Gil-Marin}, {Mueller}, {Habib}, {Heitmann}, {Ruhlmann-Kleider},
  {Padilla}, {Percival}, {Ross}, {Seo}, {Schneider}, \&
  {Zhao}}]{2020MNRAS.499.5486A}
{Avila}, S., {Gonzalez-Perez}, V., {Mohammad}, F.~G., {et~al.} 2020, \mnras,
  499, 5486, \dodoi{10.1093/mnras/staa2951}

\bibitem[{{Bailey et al.}(2023)}]{redrock2023}
{Bailey et al.} 2023, in prep.

\bibitem[{{Bianchi} \& {Percival}(2017)}]{2017MNRAS.472.1106B}
{Bianchi}, D., \& {Percival}, W.~J. 2017, \mnras, 472, 1106,
  \dodoi{10.1093/mnras/stx2053}

\bibitem[{{Boquien} {et~al.}(2019){Boquien}, {Burgarella}, {Roehlly}, {Buat},
  {Ciesla}, {Corre}, {Inoue}, \& {Salas}}]{2019A&A...622A.103B}
{Boquien}, M., {Burgarella}, D., {Roehlly}, Y., {et~al.} 2019, \aap, 622, A103,
  \dodoi{10.1051/0004-6361/201834156}

\bibitem[{{Bryan} \& {Norman}(1998)}]{1998ApJ...495...80B}
{Bryan}, G.~L., \& {Norman}, M.~L. 1998, \apj, 495, 80, \dodoi{10.1086/305262}

\bibitem[{{Chaussidon} {et~al.}(2023){Chaussidon}, {Y{\`e}che},
  {Palanque-Delabrouille}, {Alexander}, {Yang}, {Ahlen}, {Bailey}, {Brooks},
  {Cai}, {Chabanier}, {Davis}, {Dawson}, {de laMacorra}, {Dey}, {Dey},
  {Eftekharzadeh}, {Eisenstein}, {Fanning}, {Font-Ribera}, {Gazta{\~n}aga}, {A
  Gontcho}, {Gonzalez-Morales}, {Guy}, {Herrera-Alcantar}, {Honscheid},
  {Ishak}, {Jiang}, {Juneau}, {Kehoe}, {Kisner}, {Kov{\'a}cs}, {Kremin}, {Lan},
  {Landriau}, {Le Guillou}, {Levi}, {Magneville}, {Martini}, {Meisner},
  {Moustakas}, {Mu{\~n}oz-Guti{\'e}rrez}, {Myers}, {Newman}, {Nie}, {Percival},
  {Poppett}, {Prada}, {Raichoor}, {Ravoux}, {Ross}, {Schlafly}, {Schlegel},
  {Tan}, {Tarl{\'e}}, {Zhou}, {Zhou}, \& {Zou}}]{2023ApJ...944..107C}
{Chaussidon}, E., {Y{\`e}che}, C., {Palanque-Delabrouille}, N., {et~al.} 2023,
  \apj, 944, 107, \dodoi{10.3847/1538-4357/acb3c2}

\bibitem[{{Contreras} {et~al.}(2013){Contreras}, {Baugh}, {Norberg}, \&
  {Padilla}}]{2013MNRAS.432.2717C}
{Contreras}, S., {Baugh}, C.~M., {Norberg}, P., \& {Padilla}, N. 2013, \mnras,
  432, 2717, \dodoi{10.1093/mnras/stt629}

\bibitem[{{Cooper} {et~al.}(2023){Cooper}, {Koposov}, {Allende Prieto},
  {Manser}, {Kizhuprakkat}, {Myers}, {Dey}, {G{\"a}nsicke}, {Li}, {Rockosi},
  {Valluri}, {Najita}, {Deason}, {Raichoor}, {Wang}, {Ting}, {Kim}, {Carrillo},
  {Wang}, {Beraldo e Silva}, {Han}, {Ding}, {S{\'a}nchez-Conde}, {Aguilar},
  {Ahlen}, {Bailey}, {Belokurov}, {Brooks}, {Cunha}, {Dawson}, {de la Macorra},
  {Doel}, {Eisenstein}, {Fagrelius}, {Fanning}, {Font-Ribera}, {Forero-Romero},
  {Gazta{\~n}aga}, {Gontcho a Gontcho}, {Guy}, {Honscheid}, {Kehoe}, {Kisner},
  {Kremin}, {Landriau}, {Levi}, {Martini}, {Meisner}, {Miquel}, {Moustakas},
  {Nie}, {Palanque-Delabrouille}, {Percival}, {Poppett}, {Prada}, {Rehemtulla},
  {Schlafly}, {Schlegel}, {Schubnell}, {Sharples}, {Tarl{\'e}}, {Wechsler},
  {Weinberg}, {Zhou}, \& {Zou}}]{2023ApJ...947...37C}
{Cooper}, A.~P., {Koposov}, S.~E., {Allende Prieto}, C., {et~al.} 2023, \apj,
  947, 37, \dodoi{10.3847/1538-4357/acb3c0}

\bibitem[{{Davis} {et~al.}(1985){Davis}, {Efstathiou}, {Frenk}, \&
  {White}}]{1985ApJ...292..371D}
{Davis}, M., {Efstathiou}, G., {Frenk}, C.~S., \& {White}, S.~D.~M. 1985, \apj,
  292, 371, \dodoi{10.1086/163168}

\bibitem[{{Davis} \& {Peebles}(1983)}]{1983ApJ...267..465D}
{Davis}, M., \& {Peebles}, P.~J.~E. 1983, \apj, 267, 465,
  \dodoi{10.1086/160884}

\bibitem[{{DESI Collaboration} {et~al.}(2016{\natexlab{a}}){DESI
  Collaboration}, {Aghamousa}, {Aguilar}, {Ahlen}, {Alam}, {Allen}, {Allende
  Prieto}, {Annis}, {Bailey}, {Balland}, {Ballester}, {Baltay}, {Beaufore},
  {Bebek}, {Beers}, {Bell}, {Bernal}, {Besuner}, {Beutler}, {Blake}, {Bleuler},
  {Blomqvist}, {Blum}, {Bolton}, {Briceno}, {Brooks}, {Brownstein},
  {Buckley-Geer}, {Burden}, {Burtin}, {Busca}, {Cahn}, {Cai}, {Cardiel-Sas},
  {Carlberg}, {Carton}, {Casas}, {Castand er}, {Cervantes-Cota}, {Claybaugh},
  {Close}, {Coker}, {Cole}, {Comparat}, {Cooper}, {Cousinou}, {Crocce}, {Cuby},
  {Cunningham}, {Davis}, {Dawson}, {de la Macorra}, {De Vicente}, {Delubac},
  {Derwent}, {Dey}, {Dhungana}, {Ding}, {Doel}, {Duan}, {Ealet}, {Edelstein},
  {Eftekharzadeh}, {Eisenstein}, {Elliott}, {Escoffier}, {Evatt}, {Fagrelius},
  {Fan}, {Fanning}, {Farahi}, {Farihi}, {Favole}, {Feng}, {Fernandez},
  {Findlay}, {Finkbeiner}, {Fitzpatrick}, {Flaugher}, {Flender}, {Font-Ribera},
  {Forero-Romero}, {Fosalba}, {Frenk}, {Fumagalli}, {Gaensicke}, {Gallo},
  {Garcia-Bellido}, {Gaztanaga}, {Pietro Gentile Fusillo}, {Gerard},
  {Gershkovich}, {Giannantonio}, {Gillet}, {Gonzalez-de-Rivera},
  {Gonzalez-Perez}, {Gott}, {Graur}, {Gutierrez}, {Guy}, {Habib}, {Heetderks},
  {Heetderks}, {Heitmann}, {Hellwing}, {Herrera}, {Ho}, {Holland}, {Honscheid},
  {Huff}, {Hutchinson}, {Huterer}, {Hwang}, {Illa Laguna}, {Ishikawa},
  {Jacobs}, {Jeffrey}, {Jelinsky}, {Jennings}, {Jiang}, {Jimenez}, {Johnson},
  {Joyce}, {Jullo}, {Juneau}, {Kama}, {Karcher}, {Karkar}, {Kehoe}, {Kennamer},
  {Kent}, {Kilbinger}, {Kim}, {Kirkby}, {Kisner}, {Kitanidis}, {Kneib},
  {Koposov}, {Kovacs}, {Koyama}, {Kremin}, {Kron}, {Kronig}, {Kueter-Young},
  {Lacey}, {Lafever}, {Lahav}, {Lambert}, {Lampton}, {Land riau}, {Lang},
  {Lauer}, {Le Goff}, {Le Guillou}, {Le Van Suu}, {Lee}, {Lee}, {Leitner},
  {Lesser}, {Levi}, {L'Huillier}, {Li}, {Liang}, {Lin}, {Linder}, {Loebman},
  {Luki{\'c}}, {Ma}, {MacCrann}, {Magneville}, {Makarem}, {Manera}, {Manser},
  {Marshall}, {Martini}, {Massey}, {Matheson}, {McCauley}, {McDonald},
  {McGreer}, {Meisner}, {Metcalfe}, {Miller}, {Miquel}, {Moustakas}, {Myers},
  {Naik}, {Newman}, {Nichol}, {Nicola}, {Nicolati da Costa}, {Nie}, {Niz},
  {Norberg}, {Nord}, {Norman}, {Nugent}, {O'Brien}, {Oh}, {Olsen}, {Padilla},
  {Padmanabhan}, {Padmanabhan}, {Palanque-Delabrouille}, {Palmese},
  {Pappalardo}, {P{\^a}ris}, {Park}, {Patej}, {Peacock}, {Peiris}, {Peng},
  {Percival}, {Perruchot}, {Pieri}, {Pogge}, {Pollack}, {Poppett}, {Prada},
  {Prakash}, {Probst}, {Rabinowitz}, {Raichoor}, {Ree}, {Refregier}, {Regal},
  {Reid}, {Reil}, {Rezaie}, {Rockosi}, {Roe}, {Ronayette}, {Roodman}, {Ross},
  {Ross}, {Rossi}, {Rozo}, {Ruhlmann-Kleider}, {Rykoff}, {Sabiu}, {Samushia},
  {Sanchez}, {Sanchez}, {Schlegel}, {Schneider}, {Schubnell}, {Secroun},
  {Seljak}, {Seo}, {Serrano}, {Shafieloo}, {Shan}, {Sharples}, {Sholl},
  {Shourt}, {Silber}, {Silva}, {Sirk}, {Slosar}, {Smith}, {Smoot}, {Som},
  {Song}, {Sprayberry}, {Staten}, {Stefanik}, {Tarle}, {Sien Tie}, {Tinker},
  {Tojeiro}, {Valdes}, {Valenzuela}, {Valluri}, {Vargas-Magana}, {Verde},
  {Walker}, {Wang}, {Wang}, {Weaver}, {Weaverdyck}, {Wechsler}, {Weinberg},
  {White}, {Yang}, {Yeche}, {Zhang}, {Zhao}, {Zheng}, {Zhou}, {Zhou}, {Zhu},
  {Zou}, \& {Zu}}]{2016arXiv161100036D}
{DESI Collaboration}, {Aghamousa}, A., {Aguilar}, J., {et~al.}
  2016{\natexlab{a}}, arXiv e-prints, arXiv:1611.00036.
\newblock \doarXiv{1611.00036}

\bibitem[{{DESI Collaboration} {et~al.}(2016{\natexlab{b}}){DESI
  Collaboration}, {Aghamousa}, {Aguilar}, {Ahlen}, {Alam}, {Allen}, {Allende
  Prieto}, {Annis}, {Bailey}, {Balland}, {Ballester}, {Baltay}, {Beaufore},
  {Bebek}, {Beers}, {Bell}, {Bernal}, {Besuner}, {Beutler}, {Blake}, {Bleuler},
  {Blomqvist}, {Blum}, {Bolton}, {Briceno}, {Brooks}, {Brownstein},
  {Buckley-Geer}, {Burden}, {Burtin}, {Busca}, {Cahn}, {Cai}, {Cardiel-Sas},
  {Carlberg}, {Carton}, {Casas}, {Castander}, {Cervantes-Cota}, {Claybaugh},
  {Close}, {Coker}, {Cole}, {Comparat}, {Cooper}, {Cousinou}, {Crocce}, {Cuby},
  {Cunningham}, {Davis}, {Dawson}, {de la Macorra}, {De Vicente}, {Delubac},
  {Derwent}, {Dey}, {Dhungana}, {Ding}, {Doel}, {Duan}, {Ealet}, {Edelstein},
  {Eftekharzadeh}, {Eisenstein}, {Elliott}, {Escoffier}, {Evatt}, {Fagrelius},
  {Fan}, {Fanning}, {Farahi}, {Farihi}, {Favole}, {Feng}, {Fernandez},
  {Findlay}, {Finkbeiner}, {Fitzpatrick}, {Flaugher}, {Flender}, {Font-Ribera},
  {Forero-Romero}, {Fosalba}, {Frenk}, {Fumagalli}, {Gaensicke}, {Gallo},
  {Garcia-Bellido}, {Gaztanaga}, {Pietro Gentile Fusillo}, {Gerard},
  {Gershkovich}, {Giannantonio}, {Gillet}, {Gonzalez-de-Rivera},
  {Gonzalez-Perez}, {Gott}, {Graur}, {Gutierrez}, {Guy}, {Habib}, {Heetderks},
  {Heetderks}, {Heitmann}, {Hellwing}, {Herrera}, {Ho}, {Holland}, {Honscheid},
  {Huff}, {Hutchinson}, {Huterer}, {Hwang}, {Illa Laguna}, {Ishikawa},
  {Jacobs}, {Jeffrey}, {Jelinsky}, {Jennings}, {Jiang}, {Jimenez}, {Johnson},
  {Joyce}, {Jullo}, {Juneau}, {Kama}, {Karcher}, {Karkar}, {Kehoe}, {Kennamer},
  {Kent}, {Kilbinger}, {Kim}, {Kirkby}, {Kisner}, {Kitanidis}, {Kneib},
  {Koposov}, {Kovacs}, {Koyama}, {Kremin}, {Kron}, {Kronig}, {Kueter-Young},
  {Lacey}, {Lafever}, {Lahav}, {Lambert}, {Lampton}, {Landriau}, {Lang},
  {Lauer}, {Le Goff}, {Le Guillou}, {Le Van Suu}, {Lee}, {Lee}, {Leitner},
  {Lesser}, {Levi}, {L'Huillier}, {Li}, {Liang}, {Lin}, {Linder}, {Loebman},
  {Luki{\'c}}, {Ma}, {MacCrann}, {Magneville}, {Makarem}, {Manera}, {Manser},
  {Marshall}, {Martini}, {Massey}, {Matheson}, {McCauley}, {McDonald},
  {McGreer}, {Meisner}, {Metcalfe}, {Miller}, {Miquel}, {Moustakas}, {Myers},
  {Naik}, {Newman}, {Nichol}, {Nicola}, {Nicolati da Costa}, {Nie}, {Niz},
  {Norberg}, {Nord}, {Norman}, {Nugent}, {O'Brien}, {Oh}, {Olsen}, {Padilla},
  {Padmanabhan}, {Padmanabhan}, {Palanque-Delabrouille}, {Palmese},
  {Pappalardo}, {P{\^a}ris}, {Park}, {Patej}, {Peacock}, {Peiris}, {Peng},
  {Percival}, {Perruchot}, {Pieri}, {Pogge}, {Pollack}, {Poppett}, {Prada},
  {Prakash}, {Probst}, {Rabinowitz}, {Raichoor}, {Ree}, {Refregier}, {Regal},
  {Reid}, {Reil}, {Rezaie}, {Rockosi}, {Roe}, {Ronayette}, {Roodman}, {Ross},
  {Ross}, {Rossi}, {Rozo}, {Ruhlmann-Kleider}, {Rykoff}, {Sabiu}, {Samushia},
  {Sanchez}, {Sanchez}, {Schlegel}, {Schneider}, {Schubnell}, {Secroun},
  {Seljak}, {Seo}, {Serrano}, {Shafieloo}, {Shan}, {Sharples}, {Sholl},
  {Shourt}, {Silber}, {Silva}, {Sirk}, {Slosar}, {Smith}, {Smoot}, {Som},
  {Song}, {Sprayberry}, {Staten}, {Stefanik}, {Tarle}, {Sien Tie}, {Tinker},
  {Tojeiro}, {Valdes}, {Valenzuela}, {Valluri}, {Vargas-Magana}, {Verde},
  {Walker}, {Wang}, {Wang}, {Weaver}, {Weaverdyck}, {Wechsler}, {Weinberg},
  {White}, {Yang}, {Yeche}, {Zhang}, {Zhao}, {Zheng}, {Zhou}, {Zhou}, {Zhu},
  {Zou}, \& {Zu}}]{2016arXiv161100037D}
---. 2016{\natexlab{b}}, arXiv e-prints, arXiv:1611.00037.
\newblock \doarXiv{1611.00037}

\bibitem[{{DESI Collaboration} {et~al.}(2022){DESI Collaboration}, {Abareshi},
  {Aguilar}, {Ahlen}, {Alam}, {Alexander}, {Alfarsy}, {Allen}, {Allende
  Prieto}, {Alves}, {Ameel}, {Armengaud}, {Asorey}, {Aviles}, {Bailey},
  {Balaguera-Antol{\'\i}nez}, {Ballester}, {Baltay}, {Bault}, {Beltran},
  {Benavides}, {BenZvi}, {Berti}, {Besuner}, {Beutler}, {Bianchi}, {Blake},
  {Blanc}, {Blum}, {Bolton}, {Bose}, {Bramall}, {Brieden}, {Brodzeller},
  {Brooks}, {Brownewell}, {Buckley-Geer}, {Cahn}, {Cai}, {Canning}, {Capasso},
  {Carnero Rosell}, {Carton}, {Casas}, {Castander}, {Cervantes-Cota},
  {Chabanier}, {Chaussidon}, {Chuang}, {Circosta}, {Cole}, {Cooper}, {da
  Costa}, {Cousinou}, {Cuceu}, {Davis}, {Dawson}, {de la Cruz-Noriega}, {de la
  Macorra}, {de Mattia}, {Della Costa}, {Demmer}, {Derwent}, {Dey}, {Dey},
  {Dhungana}, {Ding}, {Dobson}, {Doel}, {Donald-McCann}, {Donaldson},
  {Douglass}, {Duan}, {Dunlop}, {Edelstein}, {Eftekharzadeh}, {Eisenstein},
  {Enriquez-Vargas}, {Escoffier}, {Evatt}, {Fagrelius}, {Fan}, {Fanning},
  {Fawcett}, {Ferraro}, {Ereza}, {Flaugher}, {Font-Ribera}, {Forero-Romero},
  {Frenk}, {Fromenteau}, {G{\"a}nsicke}, {Garcia-Quintero}, {Garrison},
  {Gazta{\~n}aga}, {Gerardi}, {Gil-Mar{\'\i}n}, {Gontcho a Gontcho},
  {Gonzalez-Morales}, {Gonzalez-de-Rivera}, {Gonzalez-Perez}, {Gordon},
  {Graur}, {Green}, {Grove}, {Gruen}, {Gutierrez}, {Guy}, {Hahn}, {Harris},
  {Herrera}, {Herrera-Alcantar}, {Honscheid}, {Howlett}, {Huterer},
  {Ir{\v{s}}i{\v{c}}}, {Ishak}, {Jelinsky}, {Jiang}, {Jimenez}, {Jing},
  {Joyce}, {Jullo}, {Juneau}, {Kara{\c{c}}ayl{\i}}, {Karamanis}, {Karcher},
  {Karim}, {Kehoe}, {Kent}, {Kirkby}, {Kisner}, {Kitaura}, {Koposov},
  {Kov{\'a}cs}, {Kremin}, {Krolewski}, {L'Huillier}, {Lahav}, {Lambert},
  {Lamman}, {Lan}, {Landriau}, {Lane}, {Lang}, {Lange}, {Lasker}, {Le Guillou},
  {Leauthaud}, {Le Van Suu}, {Levi}, {Li}, {Magneville}, {Manera}, {Manser},
  {Marshall}, {Martini}, {McCollam}, {McDonald}, {Meisner},
  {Mena-Fern{\'a}ndez}, {Meneses-Rizo}, {Mezcua}, {Miller}, {Miquel},
  {Montero-Camacho}, {Moon}, {Moustakas}, {Mueller}, {Mu{\~n}oz-Guti{\'e}rrez},
  {Myers}, {Nadathur}, {Najita}, {Napolitano}, {Neilsen}, {Newman}, {Nie},
  {Ning}, {Niz}, {Norberg}, {Noriega}, {O'Brien}, {Obuljen},
  {Palanque-Delabrouille}, {Palmese}, {Zhiwei}, {Pappalardo}, {PENG},
  {Percival}, {Perruchot}, {Pogge}, {Poppett}, {Porredon}, {Prada},
  {Prochaska}, {Pucha}, {P{\'e}rez-Fern{\'a}ndez}, {P{\'e}rez-R{\`a}fols},
  {Rabinowitz}, {Raichoor}, {Ramirez-Solano}, {Ram{\'\i}rez-P{\'e}rez},
  {Ravoux}, {Reil}, {Rezaie}, {Rocher}, {Rockosi}, {Roe}, {Roodman}, {Ross},
  {Rossi}, {Ruggeri}, {Ruhlmann-Kleider}, {Sabiu}, {Safonova}, {Said},
  {Saintonge}, {Salas Catonga}, {Samushia}, {Sanchez}, {Saulder}, {Schaan},
  {Schlafly}, {Schlegel}, {Schmoll}, {Scholte}, {Schubnell}, {Secroun}, {Seo},
  {Serrano}, {Sharples}, {Sholl}, {Silber}, {Silva}, {Sirk}, {Siudek}, {Smith},
  {Sprayberry}, {Staten}, {Stupak}, {Tan}, {Tarl{\'e}}, {Tie}, {Tojeiro},
  {Ure{\~n}a-L{\'o}pez}, {Valdes}, {Valenzuela}, {Valluri},
  {Vargas-Maga{\~n}a}, {Verde}, {Walther}, {Wang}, {Wang}, {Weaver},
  {Weaverdyck}, {Wechsler}, {Wilson}, {Yang}, {Yu}, {Yuan}, {Y{\`e}che},
  {Zhang}, {Zhang}, {Zhao}, {Zhou}, {Zhou}, {Zou}, {Zou}, {Zou}, {Zu}, \& {DESI
  Collaboration}}]{2022AJ....164..207D}
{DESI Collaboration}, {Abareshi}, B., {Aguilar}, J., {et~al.} 2022, \aj, 164,
  207, \dodoi{10.3847/1538-3881/ac882b}

\bibitem[{{DESI Collaboration} {et~al.}(2023{\natexlab{a}}){DESI
  Collaboration}, {Adame}, {Aguilar}, {Ahlen}, {Alam}, {Aldering}, {Alexander},
  {Alfarsy}, {Allende Prieto}, {Alvarez}, {Alves}, {Anand}, {Andrade-Oliveira},
  {Armengaud}, {Asorey}, {Avila}, {Aviles}, {Bailey},
  {Balaguera-Antol{\'\i}nez}, {Ballester}, {Baltay}, {Bault}, {Bautista},
  {Behera}, {Beltran}, {BenZvi}, {Beraldo e Silva}, {Bermejo-Climent}, {Berti},
  {Besuner}, {Beutler}, {Bianchi}, {Blake}, {Blum}, {Bolton}, {Brieden},
  {Brodzeller}, {Brooks}, {Brown}, {Buckley-Geer}, {Burtin}, {Cabayol-Garcia},
  {Cai}, {Canning}, {Cardiel-Sas}, {Carnero Rosell}, {Castander},
  {Cervantes-Cota}, {Chabanier}, {Chaussidon}, {Chaves-Montero}, {Chen},
  {Chuang}, {Claybaugh}, {Cole}, {Cooper}, {Cuceu}, {Davis}, {Dawson}, {de
  Belsunce}, {de la Cruz}, {de la Macorra}, {de Mattia}, {Demina},
  {Demirbozan}, {DeRose}, {Dey}, {Dey}, {Dhungana}, {Ding}, {Ding}, {Doel},
  {Doshi}, {Douglass}, {Edge}, {Eftekharzadeh}, {Eisenstein}, {Elliott},
  {Escoffier}, {Fagrelius}, {Fan}, {Fanning}, {Fawcett}, {Ferraro}, {Ereza},
  {Flaugher}, {Font-Ribera}, {Forero-S{\'a}nchez}, {Forero-Romero}, {Frenk},
  {G{\"a}nsicke}, {Garc{\'\i}a}, {Garc{\'\i}a-Bellido}, {Garcia-Quintero},
  {Garrison}, {Gil-Mar{\'\i}n}, {Golden-Marx}, {Gontcho}, {Gonzalez-Morales},
  {Gonzalez-Perez}, {Gordon}, {Graur}, {Green}, {Gruen}, {Guy}, {Hadzhiyska},
  {Hahn}, {Han}, {Hanif}, {Herrera-Alcantar}, {Honscheid}, {Hou}, {Howlett},
  {Huterer}, {Ir{\v{s}}i{\v{c}}}, {Ishak}, {Jana}, {Jiang}, {Jimenez}, {Jing},
  {Joudaki}, {Jullo}, {Juneau}, {Kizhuprakkat}, {Kara{\c{c}}ayl{\i}}, {Karim},
  {Kehoe}, {Kent}, {Khederlarian}, {Kim}, {Kirkby}, {Kisner}, {Kitaura},
  {Kneib}, {Koposov}, {Kov{\'a}cs}, {Kremin}, {Krolewski}, {L'Huillier},
  {Lambert}, {Lamman}, {Lan}, {Landriau}, {Lang}, {Lange}, {Lasker}, {Le
  Guillou}, {Leauthaud}, {Levi}, {Li}, {Linder}, {Lyons}, {Magneville},
  {Manera}, {Manser}, {Margala}, {Martini}, {McDonald}, {Medina},
  {Medina-Varela}, {Meisner}, {Mena-Fern{\'a}ndez}, {Meneses-Rizo}, {Mezcua},
  {Miquel}, {Montero-Camacho}, {Moon}, {Moore}, {Moustakas}, {Mueller},
  {Mundet}, {Mu{\~n}oz-Guti{\'e}rrez}, {Myers}, {Nadathur}, {Napolitano},
  {Neveux}, {Newman}, {Nie}, {Niz}, {Norberg}, {Noriega}, {Paillas},
  {Palanque-Delabrouille}, {Palmese}, {Zhiwei}, {Parkinson}, {Penmetsa},
  {Percival}, {P{\'e}rez-Fern{\'a}ndez}, {P{\'e}rez-R{\`a}fols}, {Pieri},
  {Poppett}, {Porredon}, {Prada}, {Pucha}, {Raichoor},
  {Ram{\'\i}rez-P{\'e}rez}, {Ramirez-Solano}, {Rashkovetskyi}, {Ravoux},
  {Rocher}, {Rockosi}, {Ross}, {Rossi}, {Ruggeri}, {Ruhlmann-Kleider}, {Sabiu},
  {Said}, {Saintonge}, {Samushia}, {Sanchez}, {Saulder}, {Schaan}, {Schlafly},
  {Schlegel}, {Scholte}, {Schubnell}, {Seo}, {Shafieloo}, {Sharples}, {Sheu},
  {Silber}, {Sinigaglia}, {Siudek}, {Slepian}, {Smith}, {Sprayberry},
  {Stephey}, {Su{\'a}rez-P{\'e}rez}, {Sun}, {Tan}, {Tarl{\'e}}, {Tojeiro},
  {Ure{\~n}a-L{\'o}pez}, {Vaisakh}, {Valcin}, {Valdes}, {Valluri},
  {Vargas-Maga{\~n}a}, {Variu}, {Verde}, {Walther}, {Wang}, {Wang}, {Weaver},
  {Weaverdyck}, {Wechsler}, {White}, {Xie}, {Yang}, {Y{\`e}che}, {Yu}, {Yuan},
  {Zhang}, {Zhang}, {Zhao}, {Zheng}, {Zhou}, {Zhou}, {Zou}, {Zou}, \&
  {Zu}}]{sv}
{DESI Collaboration}, {Adame}, A.~G., {Aguilar}, J., {et~al.}
  2023{\natexlab{a}}, arXiv e-prints, arXiv:2306.06307,
  \dodoi{10.48550/arXiv.2306.06307}

\bibitem[{{DESI Collaboration} {et~al.}(2023{\natexlab{b}}){DESI
  Collaboration}, {Adame}, {Aguilar}, {Ahlen}, {Alam}, {Aldering}, {Alexander},
  {Alfarsy}, {Allende Prieto}, {Alvarez}, {Alves}, {Anand}, {Andrade-Oliveira},
  {Armengaud}, {Asorey}, {Avila}, {Aviles}, {Bailey},
  {Balaguera-Antol{\'\i}nez}, {Ballester}, {Baltay}, {Bault}, {Bautista},
  {Behera}, {Beltran}, {BenZvi}, {Beraldo e Silva}, {Bermejo-Climent}, {Berti},
  {Besuner}, {Beutler}, {Bianchi}, {Blake}, {Blum}, {Bolton}, {Brieden},
  {Brodzeller}, {Brooks}, {Brown}, {Buckley-Geer}, {Burtin}, {Cabayol-Garcia},
  {Cai}, {Canning}, {Cardiel-Sas}, {Carnero Rosell}, {Castander},
  {Cervantes-Cota}, {Chabanier}, {Chaussidon}, {Chaves-Montero}, {Chen},
  {Chuang}, {Claybaugh}, {Cole}, {Cooper}, {Cuceu}, {Davis}, {Dawson}, {de
  Belsunce}, {de la Cruz}, {de la Macorra}, {de Mattia}, {Demina},
  {Demirbozan}, {DeRose}, {Dey}, {Dey}, {Dhungana}, {Ding}, {Ding}, {Doel},
  {Doshi}, {Douglass}, {Edge}, {Eftekharzadeh}, {Eisenstein}, {Elliott},
  {Escoffier}, {Fagrelius}, {Fan}, {Fanning}, {Fawcett}, {Ferraro}, {Ereza},
  {Flaugher}, {Font-Ribera}, {Forero-S{\'a}nchez}, {Forero-Romero}, {Frenk},
  {G{\"a}nsicke}, {Garc{\'\i}a}, {Garc{\'\i}a-Bellido}, {Garcia-Quintero},
  {Garrison}, {Gil-Mar{\'\i}n}, {Golden-Marx}, {Gontcho}, {Gonzalez-Morales},
  {Gonzalez-Perez}, {Gordon}, {Graur}, {Green}, {Gruen}, {Guy}, {Hadzhiyska},
  {Hahn}, {Han}, {Hanif}, {Herrera-Alcantar}, {Honscheid}, {Hou}, {Howlett},
  {Huterer}, {Ir{\v{s}}i{\v{c}}}, {Ishak}, {Jacques}, {Jana}, {Jiang},
  {Jimenez}, {Jing}, {Joudaki}, {Jullo}, {Juneau}, {Kizhuprakkat},
  {Kara{\c{c}}ayl{\i}}, {Karim}, {Kehoe}, {Kent}, {Khederlarian}, {Kim},
  {Kirkby}, {Kisner}, {Kitaura}, {Kneib}, {Koposov}, {Kov{\'a}cs}, {Kremin},
  {Krolewski}, {L'Huillier}, {Lambert}, {Lamman}, {Lan}, {Landriau}, {Lang},
  {Lange}, {Lasker}, {Le Guillou}, {Leauthaud}, {Levi}, {Li}, {Linder},
  {Lyons}, {Magneville}, {Manera}, {Manser}, {Margala}, {Martini}, {McDonald},
  {Medina}, {Medina-Varela}, {Meisner}, {Mena-Fern{\'a}ndez}, {Meneses-Rizo},
  {Mezcua}, {Miquel}, {Montero-Camacho}, {Moon}, {Moore}, {Moustakas},
  {Mueller}, {Mundet}, {Mu{\~n}oz-Guti{\'e}rrez}, {Myers}, {Nadathur},
  {Napolitano}, {Neveux}, {Newman}, {Nie}, {Nikutta}, {Niz}, {Norberg},
  {Noriega}, {Paillas}, {Palanque-Delabrouille}, {Palmese}, {Zhiwei},
  {Parkinson}, {Penmetsa}, {Percival}, {P{\'e}rez-Fern{\'a}ndez},
  {P{\'e}rez-R{\`a}fols}, {Pieri}, {Poppett}, {Porredon}, {Pothier}, {Prada},
  {Pucha}, {Raichoor}, {Ram{\'\i}rez-P{\'e}rez}, {Ramirez-Solano},
  {Rashkovetskyi}, {Ravoux}, {Rocher}, {Rockosi}, {Ross}, {Rossi}, {Ruggeri},
  {Ruhlmann-Kleider}, {Sabiu}, {Said}, {Saintonge}, {Samushia}, {Sanchez},
  {Saulder}, {Schaan}, {Schlafly}, {Schlegel}, {Scholte}, {Schubnell}, {Seo},
  {Shafieloo}, {Sharples}, {Sheu}, {Silber}, {Sinigaglia}, {Siudek}, {Slepian},
  {Smith}, {Sprayberry}, {Stephey}, {Su{\'a}rez-P{\'e}rez}, {Sun}, {Tan},
  {Tarl{\'e}}, {Tojeiro}, {Ure{\~n}a-L{\'o}pez}, {Vaisakh}, {Valcin}, {Valdes},
  {Valluri}, {Vargas-Maga{\~n}a}, {Variu}, {Verde}, {Walther}, {Wang}, {Wang},
  {Weaver}, {Weaverdyck}, {Wechsler}, {White}, {Xie}, {Yang}, {Y{\`e}che},
  {Yu}, {Yuan}, {Zhang}, {Zhang}, {Zhao}, {Zheng}, {Zhou}, {Zhou}, {Zou},
  {Zou}, \& {Zu}}]{dr}
---. 2023{\natexlab{b}}, arXiv e-prints, arXiv:2306.06308,
  \dodoi{10.48550/arXiv.2306.06308}

\bibitem[{{Dey} {et~al.}(2019){Dey}, {Schlegel}, {Lang}, {Blum}, {Burleigh},
  {Fan}, {Findlay}, {Finkbeiner}, {Herrera}, {Juneau}, {Landriau}, {Levi},
  {McGreer}, {Meisner}, {Myers}, {Moustakas}, {Nugent}, {Patej}, {Schlafly},
  {Walker}, {Valdes}, {Weaver}, {Y{\`e}che}, {Zou}, {Zhou}, {Abareshi},
  {Abbott}, {Abolfathi}, {Aguilera}, {Alam}, {Allen}, {Alvarez}, {Annis},
  {Ansarinejad}, {Aubert}, {Beechert}, {Bell}, {BenZvi}, {Beutler}, {Bielby},
  {Bolton}, {Brice{\~n}o}, {Buckley-Geer}, {Butler}, {Calamida}, {Carlberg},
  {Carter}, {Casas}, {Castander}, {Choi}, {Comparat}, {Cukanovaite}, {Delubac},
  {DeVries}, {Dey}, {Dhungana}, {Dickinson}, {Ding}, {Donaldson}, {Duan},
  {Duckworth}, {Eftekharzadeh}, {Eisenstein}, {Etourneau}, {Fagrelius},
  {Farihi}, {Fitzpatrick}, {Font-Ribera}, {Fulmer}, {G{\"a}nsicke},
  {Gaztanaga}, {George}, {Gerdes}, {Gontcho}, {Gorgoni}, {Green}, {Guy},
  {Harmer}, {Hernandez}, {Honscheid}, {Huang}, {James}, {Jannuzi}, {Jiang},
  {Joyce}, {Karcher}, {Karkar}, {Kehoe}, {Kneib}, {Kueter-Young}, {Lan},
  {Lauer}, {Le Guillou}, {Le Van Suu}, {Lee}, {Lesser}, {Perreault Levasseur},
  {Li}, {Mann}, {Marshall}, {Mart{\'\i}nez-V{\'a}zquez}, {Martini}, {du Mas des
  Bourboux}, {McManus}, {Meier}, {M{\'e}nard}, {Metcalfe},
  {Mu{\~n}oz-Guti{\'e}rrez}, {Najita}, {Napier}, {Narayan}, {Newman}, {Nie},
  {Nord}, {Norman}, {Olsen}, {Paat}, {Palanque-Delabrouille}, {Peng},
  {Poppett}, {Poremba}, {Prakash}, {Rabinowitz}, {Raichoor}, {Rezaie},
  {Robertson}, {Roe}, {Ross}, {Ross}, {Rudnick}, {Safonova}, {Saha},
  {S{\'a}nchez}, {Savary}, {Schweiker}, {Scott}, {Seo}, {Shan}, {Silva},
  {Slepian}, {Soto}, {Sprayberry}, {Staten}, {Stillman}, {Stupak}, {Summers},
  {Sien Tie}, {Tirado}, {Vargas-Maga{\~n}a}, {Vivas}, {Wechsler}, {Williams},
  {Yang}, {Yang}, {Yapici}, {Zaritsky}, {Zenteno}, {Zhang}, {Zhang}, {Zhou}, \&
  {Zhou}}]{2019AJ....157..168D}
{Dey}, A., {Schlegel}, D.~J., {Lang}, D., {et~al.} 2019, \aj, 157, 168,
  \dodoi{10.3847/1538-3881/ab089d}

\bibitem[{{Favole} {et~al.}(2017){Favole}, {Rodr{\'\i}guez-Torres}, {Comparat},
  {Prada}, {Guo}, {Klypin}, \& {Montero-Dorta}}]{2017MNRAS.472..550F}
{Favole}, G., {Rodr{\'\i}guez-Torres}, S.~A., {Comparat}, J., {et~al.} 2017,
  \mnras, 472, 550, \dodoi{10.1093/mnras/stx1980}

\bibitem[{{Favole} {et~al.}(2016){Favole}, {Comparat}, {Prada}, {Yepes},
  {Jullo}, {Niemiec}, {Kneib}, {Rodr{\'\i}guez-Torres}, {Klypin}, {Skibba},
  {McBride}, {Eisenstein}, {Schlegel}, {Nuza}, {Chuang}, {Delubac},
  {Y{\`e}che}, \& {Schneider}}]{2016MNRAS.461.3421F}
{Favole}, G., {Comparat}, J., {Prada}, F., {et~al.} 2016, \mnras, 461, 3421,
  \dodoi{10.1093/mnras/stw1483}

\bibitem[{{Foreman-Mackey}(2016)}]{2016JOSS....1...24F}
{Foreman-Mackey}, D. 2016, The Journal of Open Source Software, 1, 24,
  \dodoi{10.21105/joss.00024}

\bibitem[{{Foreman-Mackey} {et~al.}(2013){Foreman-Mackey}, {Hogg}, {Lang}, \&
  {Goodman}}]{2013PASP..125..306F}
{Foreman-Mackey}, D., {Hogg}, D.~W., {Lang}, D., \& {Goodman}, J. 2013, \pasp,
  125, 306, \dodoi{10.1086/670067}

\bibitem[{{Gao} {et~al.}(2022){Gao}, {Jing}, {Zheng}, \&
  {Xu}}]{2022ApJ...928...10G}
{Gao}, H., {Jing}, Y.~P., {Zheng}, Y., \& {Xu}, K. 2022, \apj, 928, 10,
  \dodoi{10.3847/1538-4357/ac501b}

\bibitem[{{Gao} {et~al.}(2023){Gao}, {Jing}, {Gui}, {Xu}, {Zheng}, {Zhao},
  {Aguilar}, {Ahlen}, {Brooks}, {Claybaugh}, {Dawson}, {xde la Macorra},
  {Doel}, {Fanning}, {Forero-Romero}, {A Gontcho}, {Guy}, {Honscheid}, {Kehoe},
  {Landriau}, {Manera}, {Meisner}, {Miquel}, {Moustakas}, {Newman}, {Nie},
  {Percival}, {Rossi}, {Schubnell}, {Seo}, {Tarl{\'e}}, {Weaver}, {Yu}, \&
  {Zhou}}]{Gao}
{Gao}, H., {Jing}, Y.~P., {Gui}, S., {et~al.} 2023, \apj, 954, 207,
  \dodoi{10.3847/1538-4357/ace90a}

\bibitem[{{Geach} {et~al.}(2012){Geach}, {Sobral}, {Hickox}, {Wake}, {Smail},
  {Best}, {Baugh}, \& {Stott}}]{2012MNRAS.426..679G}
{Geach}, J.~E., {Sobral}, D., {Hickox}, R.~C., {et~al.} 2012, \mnras, 426, 679,
  \dodoi{10.1111/j.1365-2966.2012.21725.x}

\bibitem[{{Gonzalez-Perez} {et~al.}(2018){Gonzalez-Perez}, {Comparat},
  {Norberg}, {Baugh}, {Contreras}, {Lacey}, {McCullagh}, {Orsi}, {Helly}, \&
  {Humphries}}]{2018MNRAS.474.4024G}
{Gonzalez-Perez}, V., {Comparat}, J., {Norberg}, P., {et~al.} 2018, \mnras,
  474, 4024, \dodoi{10.1093/mnras/stx2807}

\bibitem[{{Gonzalez-Perez} {et~al.}(2020){Gonzalez-Perez}, {Cui}, {Contreras},
  {Baugh}, {Comparat}, {Griffin}, {Helly}, {Knebe}, {Lacey}, \&
  {Norberg}}]{2020MNRAS.498.1852G}
{Gonzalez-Perez}, V., {Cui}, W., {Contreras}, S., {et~al.} 2020, \mnras, 498,
  1852, \dodoi{10.1093/mnras/staa2504}

\bibitem[{{Guo} {et~al.}(2015){Guo}, {Zheng}, {Zehavi}, {Dawson}, {Skibba},
  {Tinker}, {Weinberg}, {White}, \& {Schneider}}]{2015MNRAS.446..578G}
{Guo}, H., {Zheng}, Z., {Zehavi}, I., {et~al.} 2015, \mnras, 446, 578,
  \dodoi{10.1093/mnras/stu2120}

\bibitem[{{Guo} {et~al.}(2019){Guo}, {Yang}, {Raichoor}, {Zheng}, {Comparat},
  {Gonzalez-Perez}, {Kneib}, {Schneider}, {Bizyaev}, {Oravetz}, {Oravetz}, \&
  {Pan}}]{2019ApJ...871..147G}
{Guo}, H., {Yang}, X., {Raichoor}, A., {et~al.} 2019, \apj, 871, 147,
  \dodoi{10.3847/1538-4357/aaf9ad}

\bibitem[{{Guy} {et~al.}(2023){Guy}, {Bailey}, {Kremin}, {Alam}, {Alexander},
  {Allende Prieto}, {BenZvi}, {Bolton}, {Brooks}, {Chaussidon}, {Cooper},
  {Dawson}, {de la Macorra}, {Dey}, {Dey}, {Dhungana}, {Eisenstein},
  {Font-Ribera}, {Forero-Romero}, {Gazta{\~n}aga}, {Gontcho A Gontcho},
  {Green}, {Honscheid}, {Ishak}, {Kehoe}, {Kirkby}, {Kisner}, {Koposov}, {Lan},
  {Landriau}, {Le Guillou}, {Levi}, {Magneville}, {Manser}, {Martini},
  {Meisner}, {Miquel}, {Moustakas}, {Myers}, {Newman}, {Nie},
  {Palanque-Delabrouille}, {Percival}, {Poppett}, {Prada}, {Raichoor},
  {Ravoux}, {Ross}, {Schlafly}, {Schlegel}, {Schubnell}, {Sharples},
  {Tarl{\'e}}, {Weaver}, {Y{\'e}che}, {Zhou}, {Zhou}, \&
  {Zou}}]{2023AJ....165..144G}
{Guy}, J., {Bailey}, S., {Kremin}, A., {et~al.} 2023, \aj, 165, 144,
  \dodoi{10.3847/1538-3881/acb212}

\bibitem[{{Hadzhiyska} {et~al.}(2021){Hadzhiyska}, {Tacchella}, {Bose}, \&
  {Eisenstein}}]{2021MNRAS.502.3599H}
{Hadzhiyska}, B., {Tacchella}, S., {Bose}, S., \& {Eisenstein}, D.~J. 2021,
  \mnras, 502, 3599, \dodoi{10.1093/mnras/stab243}

\bibitem[{{Hadzhiyska} {et~al.}(2022{\natexlab{a}}){Hadzhiyska}, {Hernquist},
  {Eisenstein}, {Delgado}, {Bose}, {Kannan}, {Pakmor}, {Springel}, {Contreras},
  {Barrera}, {Ferlito}, {Hern{\'a}ndez-Aguayo}, {White}, \&
  {Frenk}}]{2022arXiv221010068H}
{Hadzhiyska}, B., {Hernquist}, L., {Eisenstein}, D., {et~al.}
  2022{\natexlab{a}}, arXiv e-prints, arXiv:2210.10068.
\newblock \doarXiv{2210.10068}

\bibitem[{{Hadzhiyska} {et~al.}(2022{\natexlab{b}}){Hadzhiyska}, {Eisenstein},
  {Hernquist}, {Pakmor}, {Bose}, {Delgado}, {Contreras}, {Kannan}, {White},
  {Springel}, {Frenk}, {Hern{\'a}ndez-Aguayo}, {Ferlito}, \&
  {Barrera}}]{2022arXiv221010072H}
{Hadzhiyska}, B., {Eisenstein}, D., {Hernquist}, L., {et~al.}
  2022{\natexlab{b}}, arXiv e-prints, arXiv:2210.10072.
\newblock \doarXiv{2210.10072}

\bibitem[{{Hahn} {et~al.}(2023){Hahn}, {Wilson}, {Ruiz-Macias}, {Cole},
  {Weinberg}, {Moustakas}, {Kremin}, {Tinker}, {Smith}, {Wechsler}, {Ahlen},
  {Alam}, {Bailey}, {Brooks}, {Cooper}, {Davis}, {Dawson}, {Dey}, {Dey},
  {Eftekharzadeh}, {Eisenstein}, {Fanning}, {Forero-Romero}, {Frenk},
  {Gazta{\~n}aga}, {Gontcho A Gontcho}, {Guy}, {Honscheid}, {Ishak}, {Juneau},
  {Kehoe}, {Kisner}, {Lan}, {Landriau}, {Le Guillou}, {Levi}, {Magneville},
  {Martini}, {Meisner}, {Myers}, {Nie}, {Norberg}, {Palanque-Delabrouille},
  {Percival}, {Poppett}, {Prada}, {Raichoor}, {Ross}, {Safonova}, {Saulder},
  {Schlafly}, {Schlegel}, {Sierra-Porta}, {Tarle}, {Weaver}, {Y{\`e}che},
  {Zarrouk}, {Zhou}, {Zhou}, \& {Zou}}]{2023AJ....165..253H}
{Hahn}, C., {Wilson}, M.~J., {Ruiz-Macias}, O., {et~al.} 2023, \aj, 165, 253,
  \dodoi{10.3847/1538-3881/accff8}

\bibitem[{{Hamilton}(1992)}]{1992ApJ...385L...5H}
{Hamilton}, A.~J.~S. 1992, \apjl, 385, L5, \dodoi{10.1086/186264}

\bibitem[{{Han} {et~al.}(2018){Han}, {Cole}, {Frenk}, {Benitez-Llambay}, \&
  {Helly}}]{2018MNRAS.474..604H}
{Han}, J., {Cole}, S., {Frenk}, C.~S., {Benitez-Llambay}, A., \& {Helly}, J.
  2018, \mnras, 474, 604, \dodoi{10.1093/mnras/stx2792}

\bibitem[{{Han} {et~al.}(2012){Han}, {Jing}, {Wang}, \&
  {Wang}}]{2012MNRAS.427.2437H}
{Han}, J., {Jing}, Y.~P., {Wang}, H., \& {Wang}, W. 2012, \mnras, 427, 2437,
  \dodoi{10.1111/j.1365-2966.2012.22111.x}

\bibitem[{{Hearin} {et~al.}(2015){Hearin}, {Watson}, \& {van den
  Bosch}}]{2015MNRAS.452.1958H}
{Hearin}, A.~P., {Watson}, D.~F., \& {van den Bosch}, F.~C. 2015, \mnras, 452,
  1958, \dodoi{10.1093/mnras/stv1358}

\bibitem[{{Hunter}(2007)}]{4160265}
{Hunter}, J.~D. 2007, Computing in Science Engineering, 9, 90

\bibitem[{{Jiang} {et~al.}(2008){Jiang}, {Jing}, {Faltenbacher}, {Lin}, \&
  {Li}}]{2008ApJ...675.1095J}
{Jiang}, C.~Y., {Jing}, Y.~P., {Faltenbacher}, A., {Lin}, W.~P., \& {Li}, C.
  2008, \apj, 675, 1095, \dodoi{10.1086/526412}

\bibitem[{{Jing}(2019)}]{2019SCPMA..6219511J}
{Jing}, Y. 2019, Science China Physics, Mechanics, and Astronomy, 62, 19511,
  \dodoi{10.1007/s11433-018-9286-x}

\bibitem[{{Jing} \& {Suto}(2002)}]{2002ApJ...574..538J}
{Jing}, Y.~P., \& {Suto}, Y. 2002, \apj, 574, 538, \dodoi{10.1086/341065}

\bibitem[{{Kauffmann} {et~al.}(2013){Kauffmann}, {Li}, {Zhang}, \&
  {Weinmann}}]{2013MNRAS.430.1447K}
{Kauffmann}, G., {Li}, C., {Zhang}, W., \& {Weinmann}, S. 2013, \mnras, 430,
  1447, \dodoi{10.1093/mnras/stt007}

\bibitem[{{Lacerna} {et~al.}(2018){Lacerna}, {Contreras}, {Gonz{\'a}lez},
  {Padilla}, \& {Gonzalez-Perez}}]{2018MNRAS.475.1177L}
{Lacerna}, I., {Contreras}, S., {Gonz{\'a}lez}, R.~E., {Padilla}, N., \&
  {Gonzalez-Perez}, V. 2018, \mnras, 475, 1177, \dodoi{10.1093/mnras/stx3253}

\bibitem[{{Lan} {et~al.}(2023){Lan}, {Tojeiro}, {Armengaud}, {Prochaska},
  {Davis}, {Alexander}, {Raichoor}, {Zhou}, {Y{\`e}che}, {Balland}, {BenZvi},
  {Berti}, {Canning}, {Carr}, {Chittenden}, {Cole}, {Cousinou}, {Dawson},
  {Dey}, {Douglass}, {Edge}, {Escoffier}, {Glanville}, {A Gontcho}, {Guy},
  {Hahn}, {Howlett}, {Hwang}, {Jiang}, {Kov{\'a}cs}, {Mezcua}, {Moore},
  {Nadathur}, {Oh}, {Parkinson}, {Rocher}, {Ross}, {Ruhlmann-Kleider}, {Sabiu},
  {Said}, {Saulder}, {Sierra-Porta}, {Weiner}, {Yu}, {Zarrouk}, {Zhang}, {Zou},
  {Ahlen}, {Bailey}, {Brooks}, {Cooper}, {de la Macorra}, {Dey}, {Dhungana},
  {Doel}, {Eftekharzadeh}, {Fanning}, {Font-Ribera}, {Garrison},
  {Gazta{\~n}aga}, {Kehoe}, {Kisner}, {Kremin}, {Landriau}, {Le Guillou},
  {Levi}, {Magneville}, {Meisner}, {Miquel}, {Moustakas}, {Myers}, {Newman},
  {Nie}, {Palanque-Delabrouille}, {Percival}, {Poppett}, {Prada}, {Schubnell},
  {Tarl{\'e}}, {Weaver}, {Zhang}, \& {Zhou}}]{2023ApJ...943...68L}
{Lan}, T.-W., {Tojeiro}, R., {Armengaud}, E., {et~al.} 2023, \apj, 943, 68,
  \dodoi{10.3847/1538-4357/aca5fa}

\bibitem[{{Landy} \& {Szalay}(1993)}]{1993ApJ...412...64L}
{Landy}, S.~D., \& {Szalay}, A.~S. 1993, \apj, 412, 64, \dodoi{10.1086/172900}

\bibitem[{{Lasker et al.}(2023)}]{ELG_Lasker}
{Lasker et al.} 2023, in prep.

\bibitem[{{Levi} {et~al.}(2013){Levi}, {Bebek}, {Beers}, {Blum}, {Cahn},
  {Eisenstein}, {Flaugher}, {Honscheid}, {Kron}, {Lahav}, {McDonald}, {Roe},
  {Schlegel}, \& {representing the DESI collaboration}}]{2013arXiv1308.0847L}
{Levi}, M., {Bebek}, C., {Beers}, T., {et~al.} 2013, arXiv e-prints,
  arXiv:1308.0847.
\newblock \doarXiv{1308.0847}

\bibitem[{{Lin} {et~al.}(2023){Lin}, {Tinker}, {Blanton}, {Guo}, {Raichoor},
  {Comparat}, \& {Brownstein}}]{2023MNRAS.519.4253L}
{Lin}, S., {Tinker}, J.~L., {Blanton}, M.~R., {et~al.} 2023, \mnras, 519, 4253,
  \dodoi{10.1093/mnras/stac2793}

\bibitem[{{Miller} {et~al.}(2023){Miller}, {Doel}, {Gutierrez}, {Besuner},
  {Brooks}, {Gallo}, {Heetderks}, {Jelinsky}, {Kent}, {Lampton}, {Levi},
  {Liang}, {Meisner}, {Sholl}, {Silber}, {Sprayberry}, {Aguilar}, {de la
  Macorra}, {Eisenstein}, {Fanning}, {Font-Ribera}, {Gaztanaga}, {Gontcho},
  {Honscheid}, {Jimenez}, {Joyce}, {Kehoe}, {Kisner}, {Kremin}, {Landriau}, {Le
  Guillou}, {Magneville}, {Martini}, {Miquel}, {Moustakas}, {Nie}, {Percival},
  {Poppett}, {Prada}, {Rossi}, {Schlegel}, {Schubnell}, {Seo}, {Sharples},
  {Tarle}, {Vargas-Magana}, \& {Zhou}}]{corrector}
{Miller}, T.~N., {Doel}, P., {Gutierrez}, G., {et~al.} 2023, arXiv e-prints,
  arXiv:2306.06310, \dodoi{10.48550/arXiv.2306.06310}

\bibitem[{{Mohammad} {et~al.}(2020){Mohammad}, {Percival}, {Seo}, {Chapman},
  {Bianchi}, {Ross}, {Zhao}, {Lang}, {Bautista}, {Brinkmann}, {Brownstein},
  {Burtin}, {Chuang}, {Dawson}, {de la Torre}, {de Mattia}, {Eftekharzadeh},
  {Fromenteau}, {Gil-Mar{\'\i}n}, {Hou}, {Mueller}, {Neveux}, {Paviot},
  {Raichoor}, {Rossi}, {Schneider}, {Tamone}, {Tinker}, {Tojeiro}, {Vargas
  Maga{\~n}a}, \& {Zhao}}]{2020MNRAS.498..128M}
{Mohammad}, F.~G., {Percival}, W.~J., {Seo}, H.-J., {et~al.} 2020, \mnras, 498,
  128, \dodoi{10.1093/mnras/staa2344}

\bibitem[{{Moster} {et~al.}(2013){Moster}, {Naab}, \&
  {White}}]{2013MNRAS.428.3121M}
{Moster}, B.~P., {Naab}, T., \& {White}, S. D.~M. 2013, \mnras, 428, 3121,
  \dodoi{10.1093/mnras/sts261}

\bibitem[{{Myers} {et~al.}(2023){Myers}, {Moustakas}, {Bailey}, {Weaver},
  {Cooper}, {Forero-Romero}, {Abolfathi}, {Alexander}, {Brooks}, {Chaussidon},
  {Chuang}, {Dawson}, {Dey}, {Dey}, {Dhungana}, {Doel}, {Fanning},
  {Gazta{\~n}aga}, {A Gontcho}, {Gonzalez-Morales}, {Hahn}, {Herrera-Alcantar},
  {Honscheid}, {Ishak}, {Karim}, {Kirkby}, {Kisner}, {Koposov}, {Kremin},
  {Lan}, {Landriau}, {Lang}, {Levi}, {Magneville}, {Napolitano}, {Martini},
  {Meisner}, {Newman}, {Palanque-Delabrouille}, {Percival}, {Poppett}, {Prada},
  {Raichoor}, {Ross}, {Schlafly}, {Schlegel}, {Schubnell}, {Tan}, {Tarle},
  {Wilson}, {Y{\`e}che}, {Zhou}, {Zhou}, \& {Zou}}]{2023AJ....165...50M}
{Myers}, A.~D., {Moustakas}, J., {Bailey}, S., {et~al.} 2023, \aj, 165, 50,
  \dodoi{10.3847/1538-3881/aca5f9}

\bibitem[{{Okumura} {et~al.}(2021){Okumura}, {Hayashi}, {Chiu}, {Lin}, {Osato},
  {Hsieh}, \& {Lin}}]{2021PASJ...73.1186O}
{Okumura}, T., {Hayashi}, M., {Chiu}, I.~N., {et~al.} 2021, \pasj, 73, 1186,
  \dodoi{10.1093/pasj/psab068}

\bibitem[{{Oliphant}(2007)}]{4160250}
{Oliphant}, T.~E. 2007, Computing in Science Engineering, 9, 10

\bibitem[{{Pahwa} \& {Paranjape}(2017)}]{2017MNRAS.470.1298P}
{Pahwa}, I., \& {Paranjape}, A. 2017, \mnras, 470, 1298,
  \dodoi{10.1093/mnras/stx1325}

\bibitem[{{Percival} \& {Bianchi}(2017)}]{2017MNRAS.472L..40P}
{Percival}, W.~J., \& {Bianchi}, D. 2017, \mnras, 472, L40,
  \dodoi{10.1093/mnrasl/slx135}

\bibitem[{{Prada} {et~al.}(2023){Prada}, {Ereza}, {Smith}, {Lasker}, {Vaisakh},
  {Kehoe}, {Dong-P{\'a}ez}, {Siudek}, {Wang}, {Alam}, {Beutler}, {Bianchi},
  {Cole}, {Dey}, {Kirkby}, {Norberg}, {Aguilar}, {Ahlen}, {Brooks},
  {Claybaugh}, {Dawson}, {de la Macorra}, {Fanning}, {Forero-Romero},
  {Gontcho}, {Hahn}, {Honscheid}, {Ishak}, {Kisner}, {Landriau}, {Manera},
  {Meisner}, {Miquel}, {Moustakas}, {Mueller}, {Nie}, {Percival}, {Poppett},
  {Rezaie}, {Rossi}, {Sanchez}, {Schubnell}, {Tarl{\'e}}, {Vargas-Maga{\~n}a},
  {Weaver}, {Yuan}, \& {Zhou}}]{overviewSHAM}
{Prada}, F., {Ereza}, J., {Smith}, A., {et~al.} 2023, arXiv e-prints,
  arXiv:2306.06315, \dodoi{10.48550/arXiv.2306.06315}

\bibitem[{{Raichoor} {et~al.}(2020){Raichoor}, {Eisenstein}, {Karim}, {Newman},
  {Moustakas}, {Brooks}, {Dawson}, {Dey}, {Duan}, {Eftekharzadeh},
  {Gazta{\~n}aga}, {Kehoe}, {Landriau}, {Lang}, {Lee}, {Levi}, {Meisner},
  {Myers}, {Palanque-Delabrouille}, {Poppett}, {Prada}, {Ross}, {Schlegel},
  {Schubnell}, {Staten}, {Tarl{\'e}}, {Tojeiro}, {Y{\`e}che}, \&
  {Zhou}}]{2020RNAAS...4..180R}
{Raichoor}, A., {Eisenstein}, D.~J., {Karim}, T., {et~al.} 2020, Research Notes
  of the American Astronomical Society, 4, 180,
  \dodoi{10.3847/2515-5172/abc078}

\bibitem[{{Raichoor} {et~al.}(2023){Raichoor}, {Moustakas}, {Newman}, {Karim},
  {Ahlen}, {Alam}, {Bailey}, {Brooks}, {Dawson}, {de la Macorra}, {de Mattia},
  {Dey}, {Dey}, {Dhungana}, {Eftekharzadeh}, {Eisenstein}, {Fanning},
  {Font-Ribera}, {Garc{\'\i}a-Bellido}, {Gazta{\~n}aga}, {A Gontcho}, {Guy},
  {Honscheid}, {Ishak}, {Kehoe}, {Kisner}, {Kremin}, {Lan}, {Landriau}, {Le
  Guillou}, {Levi}, {Magneville}, {Manera}, {Martini}, {Meisner}, {Myers},
  {Nie}, {Palanque-Delabrouille}, {Percival}, {Poppett}, {Prada}, {Ross},
  {Ruhlmann-Kleider}, {Sabiu}, {Schlafly}, {Schlegel}, {Tarl{\'e}}, {Weaver},
  {Y{\`e}che}, {Zhou}, {Zhou}, \& {Zou}}]{2023AJ....165..126R}
{Raichoor}, A., {Moustakas}, J., {Newman}, J.~A., {et~al.} 2023, \aj, 165, 126,
  \dodoi{10.3847/1538-3881/acb213}

\bibitem[{{Raichoor et al.}(2023)}]{fba}
{Raichoor et al.} 2023, in prep.

\bibitem[{{Rocher} {et~al.}(2023){Rocher}, {Ruhlmann-Kleider}, {Burtin},
  {Yuan}, {de Mattia}, {Ross}, {Aguilar}, {Ahlen}, {Alam}, {Bianchi}, {Brooks},
  {Cole}, {Dawson}, {de la Macorra}, {Doel}, {Eisenstein}, {Fanning},
  {Forero-Romero}, {Garrison}, {Gontcho A Gontcho}, {Gonzalez-Perez}, {Guy},
  {Hadzhiyska}, {Hahn}, {Honscheid}, {Kisner}, {Landriau}, {Lasker}, {E. Levi},
  {Manera}, {Meisner}, {Miquel}, {Moustakas}, {Mueller}, {Newman}, {Nie},
  {Percival}, {Poppett}, {Qin}, {Rossi}, {Samushia}, {Sanchez}, {Schlegel},
  {Schubnell}, {Seo}, {Tarl{\'e}}, {Vargas-Maga{\~n}a}, {Weaver}, {Yu},
  {Zhang}, {Zheng}, {Zhou}, \& {Zou}}]{abacusELG_Rocher}
{Rocher}, A., {Ruhlmann-Kleider}, V., {Burtin}, E., {et~al.} 2023, \jcap, 2023,
  016, \dodoi{10.1088/1475-7516/2023/10/016}

\bibitem[{{Ruiz-Macias} {et~al.}(2020){Ruiz-Macias}, {Zarrouk}, {Cole},
  {Norberg}, {Baugh}, {Brooks}, {Dey}, {Duan}, {Eftekharzadeh}, {Eisenstein},
  {Forero-Romero}, {Gazta{\~n}aga}, {Hahn}, {Kehoe}, {Landriau}, {Lang},
  {Levi}, {Lucey}, {Meisner}, {Moustakas}, {Myers}, {Palanque-Delabrouille},
  {Poppett}, {Prada}, {Raichoor}, {Schlegel}, {Schubnell}, {Tarl{\'e}},
  {Weinberg}, {Wilson}, \& {Y{\`e}che}}]{2020RNAAS...4..187R}
{Ruiz-Macias}, O., {Zarrouk}, P., {Cole}, S., {et~al.} 2020, Research Notes of
  the American Astronomical Society, 4, 187, \dodoi{10.3847/2515-5172/abc25a}

\bibitem[{{Schlafly} {et~al.}(2023){Schlafly}, {Kirkby}, {Schlegel}, {Myers},
  {Raichoor}, {Dawson}, {Aguilar}, {Allende Prieto}, {Bailey}, {BenZvi},
  {Bermejo-Climent}, {Brooks}, {de la Macorra}, {Dey}, {Doel}, {Fanning},
  {Font-Ribera}, {Forero-Romero}, {Garc{\'\i}a-Bellido}, {Gontcho}, {Guy},
  {Hahn}, {Honscheid}, {Ishak}, {Juneau}, {Kehoe}, {Kisner}, {Kremin},
  {Landriau}, {Lang}, {Lasker}, {Levi}, {Magneville}, {Manser}, {Martini},
  {Meisner}, {Miquel}, {Moustakas}, {Newman}, {Nie}, {Palanque-Delabrouille},
  {Percival}, {Poppett}, {Rockosi}, {Ross}, {Rossi}, {Tarl{\'e}}, {Weaver},
  {Y{\`e}che}, \& {Zhou}}]{ops}
{Schlafly}, E.~F., {Kirkby}, D., {Schlegel}, D.~J., {et~al.} 2023, arXiv
  e-prints, arXiv:2306.06309, \dodoi{10.48550/arXiv.2306.06309}

\bibitem[{{Schlegel et al.}(2023)}]{dr9}
{Schlegel et al.} 2023, in prep.

\bibitem[{{Silber} {et~al.}(2023){Silber}, {Fagrelius}, {Fanning}, {Schubnell},
  {Aguilar}, {Ahlen}, {Ameel}, {Ballester}, {Baltay}, {Bebek}, {Benton Beard},
  {Besuner}, {Cardiel-Sas}, {Casas}, {Castander}, {Claybaugh}, {Dobson},
  {Duan}, {Dunlop}, {Edelstein}, {Emmet}, {Elliott}, {Evatt}, {Gershkovich},
  {Guy}, {Harris}, {Heetderks}, {Heetderks}, {Honscheid}, {Illa}, {Jelinsky},
  {Jelinsky}, {Jimenez}, {Karcher}, {Kent}, {Kirkby}, {Kneib}, {Lambert},
  {Lampton}, {Leitner}, {Levi}, {McCauley}, {Meisner}, {Miller}, {Miquel},
  {Mundet}, {Poppett}, {Rabinowitz}, {Reil}, {Roman}, {Schlegel}, {Serrano},
  {Van Shourt}, {Sprayberry}, {Tarl{\'e}}, {Tie}, {Weaverdyck}, {Zhang},
  {Azzaro}, {Bailey}, {Becerril}, {Blackwell}, {Bouri}, {Brooks},
  {Buckley-Geer}, {Castro}, {Derwent}, {Dey}, {Dhungana}, {Doel}, {Eisenstein},
  {Fahim}, {Garcia-Bellido}, {Gazta{\~n}aga}, {A Gontcho}, {Gutierrez},
  {H{\"o}rler}, {Kehoe}, {Kisner}, {Kremin}, {Kronig}, {Landriau}, {Le
  Guillou}, {Martini}, {Moustakas}, {Palanque-Delabrouille}, {Peng},
  {Percival}, {Prada}, {Allende Prieto}, {de Rivera}, {Sanchez}, {Sanchez},
  {Sharples}, {Soares-Santos}, {Schlafly}, {Weaver}, {Zhou}, {Zhu}, {Zou}, \&
  {DESI Collaboration}}]{2023AJ....165....9S}
{Silber}, J.~H., {Fagrelius}, P., {Fanning}, K., {et~al.} 2023, \aj, 165, 9,
  \dodoi{10.3847/1538-3881/ac9ab1}

\bibitem[{{Sin} {et~al.}(2017){Sin}, {Lilly}, \&
  {Henriques}}]{2017MNRAS.471.1192S}
{Sin}, L. P.~T., {Lilly}, S.~J., \& {Henriques}, B. M.~B. 2017, \mnras, 471,
  1192, \dodoi{10.1093/mnras/stx1674}

\bibitem[{{Sinha} \& {Garrison}(2020)}]{2020MNRAS.491.3022S}
{Sinha}, M., \& {Garrison}, L.~H. 2020, \mnras, 491, 3022,
  \dodoi{10.1093/mnras/stz3157}

\bibitem[{{Sun} {et~al.}(2018){Sun}, {Guo}, {Wang}, {Lacey}, {Wang}, {Gao}, \&
  {Pan}}]{2018MNRAS.477.3136S}
{Sun}, S., {Guo}, Q., {Wang}, L., {et~al.} 2018, \mnras, 477, 3136,
  \dodoi{10.1093/mnras/sty832}

\bibitem[{{Szapudi} \& {Szalay}(1998)}]{1998ApJ...494L..41S}
{Szapudi}, I., \& {Szalay}, A.~S. 1998, \apjl, 494, L41, \dodoi{10.1086/311146}

\bibitem[{{Takada} {et~al.}(2014){Takada}, {Ellis}, {Chiba}, {Greene},
  {Aihara}, {Arimoto}, {Bundy}, {Cohen}, {Dor{\'e}}, {Graves}, {Gunn},
  {Heckman}, {Hirata}, {Ho}, {Kneib}, {Le F{\`e}vre}, {Lin}, {More},
  {Murayama}, {Nagao}, {Ouchi}, {Seiffert}, {Silverman}, {Sodr{\'e}},
  {Spergel}, {Strauss}, {Sugai}, {Suto}, {Takami}, \&
  {Wyse}}]{2014PASJ...66R...1T}
{Takada}, M., {Ellis}, R.~S., {Chiba}, M., {et~al.} 2014, \pasj, 66, R1,
  \dodoi{10.1093/pasj/pst019}

\bibitem[{{The Dark Energy Survey Collaboration}(2005)}]{2005astro.ph.10346T}
{The Dark Energy Survey Collaboration}. 2005, arXiv e-prints, astro.
\newblock \doarXiv{astro-ph/0510346}

\bibitem[{{Tinker} {et~al.}(2018){Tinker}, {Hahn}, {Mao}, {Wetzel}, \&
  {Conroy}}]{2018MNRAS.477..935T}
{Tinker}, J.~L., {Hahn}, C., {Mao}, Y.-Y., {Wetzel}, A.~R., \& {Conroy}, C.
  2018, \mnras, 477, 935, \dodoi{10.1093/mnras/sty666}

\bibitem[{{van der Walt} {et~al.}(2011){van der Walt}, {Colbert}, \&
  {Varoquaux}}]{5725236}
{van der Walt}, S., {Colbert}, S.~C., \& {Varoquaux}, G. 2011, Computing in
  Science Engineering, 13, 22

\bibitem[{{Wang} \& {Jing}(2010)}]{2010MNRAS.402.1796W}
{Wang}, L., \& {Jing}, Y.~P. 2010, \mnras, 402, 1796,
  \dodoi{10.1111/j.1365-2966.2009.16007.x}

\bibitem[{{Wang} {et~al.}(2006){Wang}, {Li}, {Kauffmann}, \& {De
  Lucia}}]{2006MNRAS.371..537W}
{Wang}, L., {Li}, C., {Kauffmann}, G., \& {De Lucia}, G. 2006, \mnras, 371,
  537, \dodoi{10.1111/j.1365-2966.2006.10669.x}

\bibitem[{{Wechsler} \& {Tinker}(2018)}]{2018ARA&A..56..435W}
{Wechsler}, R.~H., \& {Tinker}, J.~L. 2018, \araa, 56, 435,
  \dodoi{10.1146/annurev-astro-081817-051756}

\bibitem[{{Weinmann} {et~al.}(2006){Weinmann}, {van den Bosch}, {Yang}, \&
  {Mo}}]{2006MNRAS.366....2W}
{Weinmann}, S.~M., {van den Bosch}, F.~C., {Yang}, X., \& {Mo}, H.~J. 2006,
  \mnras, 366, 2, \dodoi{10.1111/j.1365-2966.2005.09865.x}

\bibitem[{{White} \& {Frenk}(1991)}]{1991ApJ...379...52W}
{White}, S. D.~M., \& {Frenk}, C.~S. 1991, \apj, 379, 52,
  \dodoi{10.1086/170483}

\bibitem[{{Wright} {et~al.}(2010){Wright}, {Eisenhardt}, {Mainzer}, {Ressler},
  {Cutri}, {Jarrett}, {Kirkpatrick}, {Padgett}, {McMillan}, {Skrutskie},
  {Stanford}, {Cohen}, {Walker}, {Mather}, {Leisawitz}, {Gautier}, {McLean},
  {Benford}, {Lonsdale}, {Blain}, {Mendez}, {Irace}, {Duval}, {Liu}, {Royer},
  {Heinrichsen}, {Howard}, {Shannon}, {Kendall}, {Walsh}, {Larsen}, {Cardon},
  {Schick}, {Schwalm}, {Abid}, {Fabinsky}, {Naes}, \&
  {Tsai}}]{2010AJ....140.1868W}
{Wright}, E.~L., {Eisenhardt}, P. R.~M., {Mainzer}, A.~K., {et~al.} 2010, \aj,
  140, 1868, \dodoi{10.1088/0004-6256/140/6/1868}

\bibitem[{{Xu} {et~al.}(2022){Xu}, {Zheng}, \& {Jing}}]{2022ApJ...925...31X}
{Xu}, K., {Zheng}, Y., \& {Jing}, Y. 2022, \apj, 925, 31,
  \dodoi{10.3847/1538-4357/ac38a2}

\bibitem[{{Yang} {et~al.}(2007){Yang}, {Mo}, {van den Bosch}, {Pasquali}, {Li},
  \& {Barden}}]{2007ApJ...671..153Y}
{Yang}, X., {Mo}, H.~J., {van den Bosch}, F.~C., {et~al.} 2007, \apj, 671, 153,
  \dodoi{10.1086/522027}

\bibitem[{{Yang} {et~al.}(2012){Yang}, {Mo}, {van den Bosch}, {Zhang}, \&
  {Han}}]{2012ApJ...752...41Y}
{Yang}, X., {Mo}, H.~J., {van den Bosch}, F.~C., {Zhang}, Y., \& {Han}, J.
  2012, \apj, 752, 41, \dodoi{10.1088/0004-637X/752/1/41}

\bibitem[{{Y{\`e}che} {et~al.}(2020){Y{\`e}che}, {Palanque-Delabrouille},
  {Claveau}, {Brooks}, {Chaussidon}, {Davis}, {Dawson}, {Dey}, {Duan},
  {Eftekharzadeh}, {Eisenstein}, {Gazta{\~n}aga}, {Kehoe}, {Landriau}, {Lang},
  {Levi}, {Meisner}, {Myers}, {Newman}, {Poppett}, {Prada}, {Raichoor},
  {Schlegel}, {Schubnell}, {Staten}, {Tarl{\'e}}, \&
  {Zhou}}]{2020RNAAS...4..179Y}
{Y{\`e}che}, C., {Palanque-Delabrouille}, N., {Claveau}, C.-A., {et~al.} 2020,
  Research Notes of the American Astronomical Society, 4, 179,
  \dodoi{10.3847/2515-5172/abc01a}

\bibitem[{{Yoshikawa} {et~al.}(2003){Yoshikawa}, {Jing}, \&
  {B{\"o}rner}}]{2003ApJ...590..654Y}
{Yoshikawa}, K., {Jing}, Y.~P., \& {B{\"o}rner}, G. 2003, \apj, 590, 654,
  \dodoi{10.1086/375148}

\bibitem[{{Yu} {et~al.}(2023){Yu}, {Zhao}, {Gonzalez-Perez}, {Chuang},
  {Brodzeller}, {de Mattia}, {Kneib}, {Krolewski}, {Rocher}, {Ross}, {Wang},
  {Yuan}, {Zhang}, {Zhou}, {Aguilar}, {Ahlen}, {Brooks}, {Dawson}, {de la
  Macorra}, {Doel}, {Fanning}, {Font-Ribera}, {Forero-Romero}, {Gontcho},
  {Honscheid}, {Kehoe}, {Kisner}, {Kremin}, {Landriau}, {Manera}, {Martini},
  {Meisner}, {Miquel}, {Moustakas}, {Nie}, {Percival}, {Poppett}, {Raichoor},
  {Rossi}, {Seo}, {Tarl{\'e}}, {Zhou}, \& {Zou}}]{inclusiveSHAM}
{Yu}, J., {Zhao}, C., {Gonzalez-Perez}, V., {et~al.} 2023, arXiv e-prints,
  arXiv:2306.06313, \dodoi{10.48550/arXiv.2306.06313}

\bibitem[{{Yuan} {et~al.}(2022{\natexlab{a}}){Yuan}, {Garrison}, {Hadzhiyska},
  {Bose}, \& {Eisenstein}}]{2022MNRAS.510.3301Y}
{Yuan}, S., {Garrison}, L.~H., {Hadzhiyska}, B., {Bose}, S., \& {Eisenstein},
  D.~J. 2022{\natexlab{a}}, \mnras, 510, 3301, \dodoi{10.1093/mnras/stab3355}

\bibitem[{{Yuan} {et~al.}(2022{\natexlab{b}}){Yuan}, {Hadzhiyska}, {Bose}, \&
  {Eisenstein}}]{2022MNRAS.512.5793Y}
{Yuan}, S., {Hadzhiyska}, B., {Bose}, S., \& {Eisenstein}, D.~J.
  2022{\natexlab{b}}, \mnras, 512, 5793, \dodoi{10.1093/mnras/stac830}

\bibitem[{{Yuan} {et~al.}(2023{\natexlab{a}}){Yuan}, {Zhang}, {Ross},
  {Donald-McCann}, {Hadzhiyska}, {Wechsler}, {Zheng}, {Alam}, {Gonzalez-Perez},
  {Aguilar}, {Ahlen}, {Bianchi}, {Brooks}, {de la Macorra}, {Fanning},
  {Forero-Romero}, {Honscheid}, {Ishak}, {Kehoe}, {Lasker}, {Landriau},
  {Manera}, {Martini}, {Meisner}, {Miquel}, {Moustakas}, {Nadathur}, {Newman},
  {Nie}, {Percival}, {Poppett}, {Rocher}, {Rossi}, {Sanchez}, {Samushia},
  {Schubnell}, {Seo}, {Tarle}, {Weaver}, {Yu}, {Zhou}, \&
  {Zou}}]{abacusLRGQSO_Yuan}
{Yuan}, S., {Zhang}, H., {Ross}, A.~J., {et~al.} 2023{\natexlab{a}}, arXiv
  e-prints, arXiv:2306.06314, \dodoi{10.48550/arXiv.2306.06314}

\bibitem[{{Yuan} {et~al.}(2023{\natexlab{b}}){Yuan}, {Wechsler}, {Wang}, {de
  los Reyes}, {Myles}, {Rocher}, {Hadzhiyska}, {Aguilar}, {Ahlen}, {Brooks},
  {Claybaugh}, {Cole}, {de la Macorra}, {Forero-Romero}, {Gontcho}, {Guy},
  {Honscheid}, {Kisner}, {Levi}, {Manera}, {Meisner}, {Miquel}, {Moustakas},
  {Nie}, {Palanque-Delabrouille}, {Poppett}, {Rezaie}, {Ross}, {Rossi},
  {Sanchez}, {Schubnel}, {Seo}, {Tarl{\'e}}, {Weaver}, \&
  {Zhou}}]{2023arXiv231009329Y}
{Yuan}, S., {Wechsler}, R.~H., {Wang}, Y., {et~al.} 2023{\natexlab{b}}, arXiv
  e-prints, arXiv:2310.09329, \dodoi{10.48550/arXiv.2310.09329}

\bibitem[{{Zhai} {et~al.}(2021){Zhai}, {Wang}, {Benson}, {Chuang}, \&
  {Yepes}}]{2021MNRAS.505.2784Z}
{Zhai}, Z., {Wang}, Y., {Benson}, A., {Chuang}, C.-H., \& {Yepes}, G. 2021,
  \mnras, 505, 2784, \dodoi{10.1093/mnras/stab1539}

\bibitem[{{Zhou} {et~al.}(2020){Zhou}, {Newman}, {Dawson}, {Eisenstein},
  {Brooks}, {Dey}, {Dey}, {Duan}, {Eftekharzadeh}, {Gazta{\~n}aga}, {Kehoe},
  {Landriau}, {Levi}, {Licquia}, {Meisner}, {Moustakas}, {Myers},
  {Palanque-Delabrouille}, {Poppett}, {Prada}, {Raichoor}, {Schlegel},
  {Schubnell}, {Staten}, {Tarl{\'e}}, \& {Y{\`e}che}}]{2020RNAAS...4..181Z}
{Zhou}, R., {Newman}, J.~A., {Dawson}, K.~S., {et~al.} 2020, Research Notes of
  the American Astronomical Society, 4, 181, \dodoi{10.3847/2515-5172/abc0f4}

\bibitem[{{Zhou} {et~al.}(2023){Zhou}, {Dey}, {Newman}, {Eisenstein}, {Dawson},
  {Bailey}, {Berti}, {Guy}, {Lan}, {Zou}, {Aguilar}, {Ahlen}, {Alam}, {Brooks},
  {de la Macorra}, {Dey}, {Dhungana}, {Fanning}, {Font-Ribera}, {Gontcho},
  {Honscheid}, {Ishak}, {Kisner}, {Kov{\'a}cs}, {Kremin}, {Landriau}, {Levi},
  {Magneville}, {Manera}, {Martini}, {Meisner}, {Miquel}, {Moustakas}, {Myers},
  {Nie}, {Palanque-Delabrouille}, {Percival}, {Poppett}, {Prada}, {Raichoor},
  {Ross}, {Schlafly}, {Schlegel}, {Schubnell}, {Tarl{\'e}}, {Weaver},
  {Wechsler}, {Y{\'e}che}, \& {Zhou}}]{2023AJ....165...58Z}
{Zhou}, R., {Dey}, B., {Newman}, J.~A., {et~al.} 2023, \aj, 165, 58,
  \dodoi{10.3847/1538-3881/aca5fb}

\bibitem[{{Zou} {et~al.}(2017){Zou}, {Zhou}, {Fan}, {Zhang}, {Zhou}, {Nie},
  {Peng}, {McGreer}, {Jiang}, {Dey}, {Fan}, {He}, {Jiang}, {Lang}, {Lesser},
  {Ma}, {Mao}, {Schlegel}, \& {Wang}}]{2017PASP..129f4101Z}
{Zou}, H., {Zhou}, X., {Fan}, X., {et~al.} 2017, \pasp, 129, 064101,
  \dodoi{10.1088/1538-3873/aa65ba}

\bibitem[{{Zu} \& {Mandelbaum}(2018)}]{2018MNRAS.476.1637Z}
{Zu}, Y., \& {Mandelbaum}, R. 2018, \mnras, 476, 1637,
  \dodoi{10.1093/mnras/sty279}

\bibitem[{{Zu} {et~al.}(2022){Zu}, {Song}, {Shao}, {Chen}, {Zheng}, {Gao},
  {Yu}, {Shan}, \& {Jing}}]{2022MNRAS.511.1789Z}
{Zu}, Y., {Song}, Y., {Shao}, Z., {et~al.} 2022, \mnras, 511, 1789,
  \dodoi{10.1093/mnras/stac125}

\end{thebibliography}
\bibliographystyle{aasjournal}

\end{document}